\documentstyle[aps,pre]{revtex} 
\input epsf.sty 
\begin{document} 
\twocolumn[ 
\hsize\textwidth\columnwidth\hsize\csname@twocolumnfalse\endcsname 
\draft

\title{Gauged Neural Network: Phase Structure, Learning, and Associative 
Memory} 
\author{Motohiro Kemuriyama and
Tetsuo Matsui}
\address{
Department of Physics, Kinki University, 
Higashi-Osaka, 577-8502 Japan}
\author{Kazuhiko Sakakibara} 
\address{
Department of Physics, Nara National College of Technology, 
Yamatokohriyama, 639-1080 Japan,
} 

  
\maketitle  

\begin{abstract}   
A gauge model of neural network is introduced, which resembles
the Z(2)  Higgs lattice gauge theory of high-energy physics. 
It contains a neuron variable $S_x = \pm 1$ on each site $x$
of a 3D lattice and a synaptic-connection variable $J_{x\mu} = \pm 1$ 
on each link $(x,x+\hat{\mu}) (\mu=1,2,3)$. 
The model is regarded as a generalization of the Hopfield model
of associative memory to a  model of learning by 
converting the synaptic weight    between $x$ and $x+\hat{\mu}$ 
to a dynamical Z(2) gauge variable  $J_{x\mu}$. 
The  local Z(2) gauge symmetry is inherited from the Hopfield model 
and  assures us the locality of time evolutions 
of $S_x$ and $J_{x\mu}$ and a generalized Hebbian 
learning rule.
At finite ``temperatures", numerical simulations show that 
the model exhibits the
Higgs, confinement, and Coulomb phases. 
We simulate dynamical processes of learning a pattern of $S_x$
and recalling it,
and classify the parameter space according to the
performance.
At some parameter regions,  stable column-layer structures 
in signal propagations 
are spontaneously generated.
Mutual interactions
between $S_x$ and $J_{x\mu}$ induce partial memory loss 
as expected.
\vspace{1cm}
\end{abstract} 

]


\setcounter{section}{0}

\setcounter{footnote}{0} 
\vspace{1cm}
\section{Introduction}

Our brains exhibit various complicated functions like recognition, 
thinking, etc. In particular, it is quite interesting to understand
the mechanism of storing and restoring particular concepts, i.e., 
how to learn and recall them.
It is challenging to construct a mathematical model 
to describe the essence of learning and recalling. 

In the framework of neural network, it is well known that the Hopfield 
model\cite{hopfield} of associative memory
offers us a reasonable mechanism how to recall the images and patterns 
that one has once learned. 
In the Hopfield model, the state of the $i$-th neuron, 
excited or unexcited, is described
by the Z(2) variable $S_i ( = \pm 1) (i=1,2,...,N)$, 
and the state of the synaptic 
connection between the $i$-th and $j$-th neurons is expressed
by its strength (weight), $J_{ij}$, to transmit signals,  
which is a preassigned real constant.  
The signal at the $j$-th site at time $t$ 
propagates to the $i$-th site through the axon and 
 the synaptic connection in the form $J_{ij}S_j(t)$ to affect
the state $S_i$ in the next time step $t +  \epsilon$ as
\begin{eqnarray}
S_i(t + \epsilon) = {\rm sgn}\left[\sum_{j=1}^NJ_{ij}S_j(t)\right].
\label{timeevolution}
\end{eqnarray}
Physically, $S_j$ is the signature of the electric potential 
at the $j$-th neuron (membrane potential) measured from a certain
common level, and $J_{ij}$ is the conversion factor to propagate 
the potential from $j$ to $i$.
So the quantity $\sum_j J_{ij} S_j$ in the bracket
is just the total amount of potential accumulated at the $i$-th 
neuron. The case $J_{ij} > 0$ enhances the tendency that
the $i$-th neuron shall take the same state as that at $j$,
$S_i =  S_j$, while the case $J_{ij} < 0$ suppresses it,
favoring the opposite state, $S_i =  -S_j$.

The time evolution (\ref{timeevolution}), when applied 
to all the variables $S_i$ in an asynchronous manner,
is known to decrease (not increase) the following ``energy", 
\begin{eqnarray}
E &=& -\frac{1}{2} \sum_{i,j=1}^N S_{i} J_{ij} S_j.
\label{hopfieldenergy}
\end{eqnarray}
Let us prepare $M$ patterns $S_i = \xi^\alpha_i$ $(\alpha = 1,2,...,M)$
(which are mutually orthogonal, $\sum_i \xi^\alpha_i \xi^{\beta}_i 
= N\delta_{\alpha \beta}$) as the patterns to learn.
 One may choose $J_{ij}$ so that 
each of these patterns,  $S_i = \xi^\alpha_i$,
is a local minimum of this energy. Such an idea is realized by
taking the following Hebbian learning rule \cite{hebb};
\begin{eqnarray}
J_{ij} &=& \frac{1}{M}\sum_{\alpha=1}^M \xi_i^\alpha \xi_j^\alpha.
\label{hebb}
\end{eqnarray}
If we start from a pattern of $S_i(0)$, 
then one of the  memorized patterns that locates nearest to 
$S_i(0)$, say $\xi^\alpha_i$, is obtained gradually via 
the rule (\ref{timeevolution}), $S_i(t) 
= \xi^\alpha_i$ in a certain time\cite{unwelcomeminima}.
This certainly offers us a possible mechanism of associative 
memory. 
 
How about processes of learning patterns?
The Hopfield model itself is incapable for this purpose, since
the memorized patterns are stored in the fixed parameters $J_{ij}$
as (\ref{hebb})
from the beginning. We need some generalization so that these $J_{ij}$
change as time goes by, i.e., plasticity of $J_{ij}$. 
A simple example of the 
time dependence of $J_{ij}$ is the following;
\begin{eqnarray}
J_{ij}(t+\epsilon) &=& J_{ij}(t) +
\frac{1}{M}\xi_i^\alpha \xi_j^\alpha.
\label{hebbJ}
\end{eqnarray}
It reflects the Hebbian rule (\ref{hebb}),
describing the process that 
the pattern $S_i=\xi_i^\alpha$ is learned
in the period $(t,t+\epsilon)$.

A well-known model of learning is a perceptron
or its improvement, the error back-propagation 
model\cite{perceptron}. In a perceptron,
one assigns a pattern $S_i = \xi_i$ for neurons $S_i$ in the input layer
and optimizes the weight $w_{i}$ connecting $S_i$ and
the single output neuron $S'$ so that $S' \equiv 
\theta(\sum_i w_i S_i) $ produces 1(0) for 
$S_i=\xi_i(S_i \neq \xi_i)$ approximately. The model proves itself 
very useful for the problems of pattern recognition.
However, in a perceptron, (i) the flow of informations is one way
from the input layer to the output neuron,
and  (ii) the time evolution of  $S_i(t)$ is missing,
although the  process of optimization  of $w_i$ itself 
may be regarded as a time evolution of $J_{ij}$.
These two points seem quite different from what is going on in 
our brains. 

In a set of models of 
self-organization\cite{selforganization,malsburg,tanaka,linsker},
signals from the neurons $S_j$ in the input layer are forwarded
with the weights $w_{ij}$ to the neurons $S'_i$ in the output layer,
and $w_{ij}(t)$ are assumed to evolve with the rule like 
$w_{ij}(t+\epsilon)-w_{ij}(t) = -\alpha w_{ij} + \beta S'_i S_j$, 
i.e., with the damping effect 
 and the Hebbian rule (and/or certain constraints on $w_{ij}$). 
 It is shown that the neurons in the output layer 
are partitioned into subsets, and the neurons in each subset become 
active for a particular pattern in the input layer. This may explain
the emergence of so-called column structures of active neurons 
 observed in human brains
like ocular dominance columns, orientation columns, and direction 
columns in the medial temporal area.
However, these models also share the properties (i) and (ii) above.

As a stochastic model of learning, Boltzmann machines\cite{boltzmann} 
are well known. Here $S_i$ develops according to Boltzmann
distribution for the energy (\^ref{hopfieldenergy}) 
and the learning rule is like
$w_{ij}(t+\epsilon)-w_{ij}(t) = \beta (\langle S_i S_j \rangle_\xi
-\langle S_i S_j \rangle_{E})$, where the first term in R.H.S. 
is an average over patterns to learn and describes the Hebbian rule
whereas the second term is an average over Bolotzmann  distribution
and describes so-called unlearning processes. 

Following these pioneering works,
various models of learning have appeared. 
Some models
treat both $S_i (=\pm1)$ and $J_{ij} [\in (-\infty, \infty)]$
as dynamical variables\cite{suzuki}.
Although they contain certain interesting features,
it seems necessary to consider some natural and explicit principles 
that characterize the dynamics of $S_i$ and $J_{ij}$.

In Ref.\cite{matsui}, a gauge model of neural network is proposed.
Here $J_{ij}$ are regarded as gauge variables, i.e., 
path-dependent phase factors (exponentiated gauge field) and
the energy of the model is obtained by adding
a couple of additional gauge-invariant terms 
to the Hopfield energy (\ref{hopfieldenergy}). 
Both $S_i$ and $J_{ij}$ are treated as dynamical variables
and their time-dependences at finite ``temperature" ($T$)
are postulated to obey a stochastic process for Boltzmann distribution
of the gauge-invariant  energy. 
The model on a 3D lattice at finite $T$ 
resembles the lattice gauge theory\cite{wilson} in 
high-energy particle physics. The  phase structure of lattice gauge 
theory was intensively examined by mean field theory (MFT) and 
Monte Carlo (MC) simulations\cite{lgt}.  Generally, three phases,
Higgs phase, Coulomb phase, and confinement phase, are known to
be possible [see the table 
(\ref{tab:phases}) below]. In the gauge model of neural network,
the former two correspond to the ferromagnetic phase
($\langle J_{ij} \rangle  \neq 0,\ \langle S_i \rangle \neq 0$ in MFT) 
and the paramagnetic phase
($ \langle J_{ij} \rangle \neq 0,\ \langle S_i \rangle = 0$) 
in the Hopfield model, respectively.
The third confinement phase has $\langle J_{ij} \rangle  = 0$ and
$\langle S_i \rangle =0$, describing the state 
that both learning and recalling are disable.
For example, the Higgs phase corresponds to a smart student
who took a lesson  and made a good score in examination;
the Coulomb phase to a student who took a lesson but with
a bad score; the confinement phase to a student who cuts a class.

There is  a quite  interesting relation between
the present gauge model and the quantum memory of a 
toric code studied by Kitaev et al.\cite{kitaev}.
To calculate the accuracy threshold of a 2D toric quantum memory,
one needs to study the phase diagram of the  3D pure Z(2)
lattice gauge theory with random gauge coupling.
The  present model at $c_1=c_3=0$ (see Sect.IIIA) coincides
this gauge model at vanishing randomness.
The error-free condition in  uploading data to
and  downloading from a quantum memory leads us  that
this pure Z(2) gauge model should be in the ordered (Coulomb) 
phase\cite{kitaev2}. 
Thus the ability of leaning patterns
 in the gauged neural network
corresponds to the error-free function of a
 quantum memory of a toric code.

In this paper, we study various aspects of this Z(2) 
gauged neural network in detail.
We investigate the full phase structure of the system 
at finite  $T$. We   
also simulate the dynamical processes of learning a pattern of $S_i$
and recalling it,
and examine the parameter dependence of the rate of performance. 

Here we note that Thoulouse\cite{toulouse} noticed Z(2) 
gauge symmetry in  the related models, the  models 
of spin glasses\cite{sg},  
and adopted it to characterize effects of frustration.
In fact, the Sherrington-Kirkpatrick model of 
spin glasses\cite{sg} has the same form of energy as that of 
the Hopfield model (\ref{hopfieldenergy}). 
Since then, there have appeared various studies on frustrations
in spin glasses\cite{toulouse2}.
The crucial difference between the models of spin glasses and
the present model of neural network is in that the 
``gauge variables" $J_{ij}$
are regarded as quenched random variables in the spin-glass theory
whereas they are treated in the  present neural network 
as dynamical variables
on an equal footing to the dynamical neuron variables $S_i$.

The structure of the paper is as follows; 
In Sect.II, we explain the origin and the relevance of 
Z(2) gauge symmetry in neural networks in details.
In Sect.III, we introduce the Z(2) gauge model on a 3D lattice.
In Sect.IV, we study its phase structure at finite $T$ 
by statistical mechanics and find  emergent 
column-layer structures. In Sect.V, we simulate dynamical 
processes of learning a pattern and recalling it. 
In Sect.VI, we present discussion and the problems in future.

\section{Z(2) Gauge Symmetry}
\setcounter{equation}{0}

In this section, we explain the origin of the gauge symmetry 
and its relevance to neural networks.

Let us start with the Hopfield model (\ref{hopfieldenergy}).
Here  one may assign  $S_i = 1$ 
for the excited (fired) state of the $i$-th neuron
and $S_i = -1$ for the unexcited state.
Let us focus on the part of network consisting of the $j$-th neuron 
and the axon and synaptic connection starting from the $j$-th neuron 
and ending at the $i$-th neuron, and consider the following 
 two states (a) and (b) of this part;
\begin{eqnarray}
&{\rm state\ (a):}\ \ \ \ & (S_j, J_{ij}), \nonumber\\
& {\rm state\ (b):}\ \ \ \ &  (S_j' = -S_j, J_{ij}' = -J_{ij}).
\label{statesab}
\end{eqnarray} 
Physically, they are independent(different) each other.
For example, (a)$\ (S_j=1, J_{ij}=J>0)$ describes the excited neuron and 
the enhancing connection, while  (b)$\ (S_j=-1, J_{ij} = -J < 0)$
describes the unexcited neuron and the suppressing (inhibitory) connection.
In the Hopfield model, these two states (a) and (b) are degenerate; 
they have the same amount of energy (\ref{hopfieldenergy}).
Actually,  the energy $E_{ij}$ stored in the above part is
\begin{eqnarray}
E_{ij}(b) &\equiv& - S_{i} J'_{ij} S'_j 
= -S_{i} (-J_{ij})(-S_j) \nonumber\\
&=& -S_{i} J_{ij} S_j = E_{ij}(a).
\label{Eij}
\end{eqnarray}

The state (b) is obtained from (a) by the following 
replacement of $S_j$ and $J_{ij}$;
\begin{eqnarray}
S_j &\rightarrow & S_j'\equiv  V_j S_j,\ \ J_{ij} \rightarrow 
J_{ij}' \equiv J_{ij}V_j,\ \ V_j = -1.
\label{z2transform}
\end{eqnarray}
In lattice gauge theory\cite{wilson}, the replacement (\ref{z2transform})
is known as a local Z(2) gauge transformation in which $J_{ij}$
plays the role of a so-called gauge variable.
More generally, one may consider the following general local
(site-dependent) Z(2) gauge transformation throughout 
the network (for all $i$ and $j$);
\begin{eqnarray}
S_i &\rightarrow & S_i'\equiv  V_i S_i,\ \ J_{ij} \rightarrow 
J_{ij}' \equiv V_iJ_{ij}V_j,\ \ V_i = \pm1.
\label{z2transform2}
\end{eqnarray}
Here, whether one replaces the state (a) by (b) ($V_i = -1$)
or not ($V_i = 1$) may depend site by site. 
It is easy to see that the Hopfield energy (\ref{hopfieldenergy}) 
is invariant under this transformation. 

We regard this ``gauge symmetry" of the Hopfield model as an
important property that the  generalized model of learning
should inherit.
To generalize the Hopfield model to such a gauge model of learning,  
we just need  (i)  to convert  $J_{ij}$ from constants to
dynamical (gauge) variables, 
which are to transform as (\ref{z2transform2}), 
and (ii) to introduce a gauge-invariant energy.

In  Fig.\ref{fig:gauge} we illustrate the function of 
the gauge field $A_{\mu}(x)$ 
in the conventional U(1) gauge field theory
by using the path-dependent phase factor,  
\begin{eqnarray} 
U(x,y) \equiv  \exp\left(i\int_{P_{xy}} dx'_\mu A_{\mu}(x')\right),
\end{eqnarray}
where $P_{xy}$ is a certain path connecting two spatial points $y$ and $x$.
It conveys informations about the relative orientation of
two internal coordinates at $x$ and $y$,
and parallel-translates a vector 
$\varphi(y)$ at $y$ to $x$ along a path from $y$ to $x$, 
giving rise to a vector $U(x,y)\varphi(y)$ at $x$.

Since the strength $J_{ij}$ of the synaptic connection 
 describes the way how the electric signal at $j$ is transformed to $i$,
it is natural also from this gauge-theoretical point of view
to identify $J_{ij}$ as $U(x,y)$
in gauge theory.
Then the resulting signal conveyed from $j$ to $i$, 
$J_{ij}S_j$, corresponds to $U(x,y)\varphi(y)$.

Let us explain the relation between
the gauge symmetry and the physical states in detail\cite{yangmills}.
One may first think that the assignment  of
the gauge-variant variable $S_i=\pm1$ to the 
the physical (excited and unexcited) states is 
inappropriate, because a gauge transformation $S_i' = -S_i$ 
exchanges the excited state and the unexcited state.
However, such an  assignment brings no difficulties
because $J_{ij}$ also changes. 
Let us start with the global frame of $S_i$ illustrated in
Fig.\ref{fig:localframe}a, where
$S_i=1(-1)$ describe the excited(unexcited) 
state of the $i$-th neuron for all $i$. 
Suppose that we make a local gauge transformation with $V_i = 1-2\delta_{ij}$, which 
implies that $S'_j=1(-1)$ describes the unexcited(excited) state.  
See Fig.\ref{fig:localframe}b.
Such a capricious or perverse choice   is ``corrected" or ``compensated" 
by the associated transformation 
$J'_{ij} = -J_{ij}$, thus bringing no changes in physical content.
Actually, the state at $j$ is to be recognized at the neighboring
observer at $i$ by the combination
$J_{ij}S_j$. There holds the equality $J_{ij}S_j = J'_{ij}S'_j$,
which shows that the physical state at $j$ does not change.

From these considerations, we regard  $J_{ij}$ as the path-dependent 
phase factors of the gauge group Z(2), hence they take
Z(2) variables, 
\begin{eqnarray}
J_{ij} = \pm1 \in Z(2)
\label{jijz2}
\end{eqnarray}
in appropriate unit (which is to be supplied through 
the coefficients in the energy). 
One may claim that $J_{ij}$ should take continuous real values
$-\infty < J_{ij} < \infty$ instead of Z(2) variables.
Such an assignment is certainly possible in constructing
a gauge model, as long as  $J_{ij}$ transform as (\ref{z2transform2}).
However, there should be certain constraints on  $J_{ij}$,
which reflect, e.g., conservation laws of chemical materials
and restrict their possible values.
On this point, study of the biological models of 
self-organization \cite{selforganization,malsburg,tanaka,linsker} 
is suggestive and can be 
taken as a support to our simplification (\ref{jijz2}).
For example, von der Malsburg\cite{malsburg} 
puts the constraint $\sum_j J_{ij}=1$ for each $i$,
while Tanaka et al.\cite{tanaka} argue that
the values of $J_{ij}$ can be well approximated by the Potts spin
variables; $J_{ij} = 0\ {\rm or}\ 1, \sum_{j}J_{ij}=1$. 
While both authors focus on the enhancing connections $(J_{ij} > 0)$,
Linsker\cite{linsker} considers also the case of suppressing 
connections $(J_{ij} < 0)$ and puts the constraints 
$-1 < J_{ij} < 1$.
Our treatment (\ref{jijz2}) may be viewed  
(i) as the special case of Linsker's treatment that 
most configurations are to reach the end 
points of his inequality, and/or (ii) as a 
generalization of Tanaka's result to the Z(2) gauge theory 
in which both $J_{ij}> 0$ and $J_{ij} < 0$ should appear 
as (\ref{z2transform2}) indicates\cite{jij0}. 

Concerning to the signature of $J_{ij}$.
it was once thought that each neuron releases only one 
type of neurotransmitter (Dale's law), which implies that
the signature of $J_{ij}$ should be fixed for each $j$. However,
there is now  evidence that Dale's law does not hold, indicating
neurons release more than one kind of neurotransmitter. 
Irrespective of experimental circumstances including this case, 
we have a sound reason to 
consider  $J_{ij}$ with indefinite signature.
It is supplied by regarding the gauge model under consideration 
as an effective model of renormalization-group theory 
as explained in Sect.IIIA for the explicit lattice model.
Each $S_i$ is associated there not with a single neuron
but with a cluster
of neurons, expressing the ``average" over the states of
neurons contained in each cluster. 
$J_{ij}$ is also the average strength between 
two clusters, so its signature is to be indefinite. 


As stated before, the Z(2) gauge symmetry requires that the
energy $E(\{S_i\}, \{J_{ij}\})$ of the system is gauge-invariant;
\begin{eqnarray}
S_i  (=\pm1) &\rightarrow& S'_i \equiv V_i S_i,\nonumber\\ 
J_{ij}(=\pm1) &\rightarrow& J'_{ij} \equiv V_iJ_{ij}V_j, 
\ \ V_i = \pm1,\nonumber\\
E(\{S'_i\}, \{J'_{ij}\}) &=& E(\{S_i\}, \{J_{ij}\}).
\label{gaugetransformation2}
\end{eqnarray}
The principle of 
gauge symmetry puts severe restrictions on the form of $E$. 
Some possible gauge-invariant terms are depicted 
in Fig.\ref{fig:energy0}.
The first term $S_iJ_{ij}S_j$ in Fig.\ref{fig:energy0}
 is just the term
  of the Hopfield model. The second term $S_iJ_{ik}J_{kj}S_j$
  may be viewed to describe the combined effect
  of the two successive processes
  $S_k J_{kj} S_j$ and $S_i J_{ik} S_k$.
  Actually, due to $S_k^2 = 1$, we have 
\begin{eqnarray}
&&S_i J_{ik} S_k \times S_k J_{kj} S_j =
S_i J_{ik}J_{kj} S_j. 
\label{indirect}
\end{eqnarray}  
In the same way,
the last two terms are interpreted as 
\begin{eqnarray}
J_{ij}J_{ji} &=& S_iJ_{ij}S_j \times S_j J_{ji} S_i,\nonumber\\
J_{ij}J_{jk}J_{ki}&=&S_i J_{ij} S_j \times S_j J_{jk} S_k 
\times S_k J_{ki} S_i.
\label{reverberating}
\end{eqnarray} 
These two terms describe the combined processes
taking place along closed circuits. Hebb\cite{hebb} has once 
introduced the concept of so called reverberating circuits, 
in which short-term memories are considered to be stored. These terms
(\ref{reverberating}) may describe functions of such circuits.
%

\section{Model}

\subsection{Z(2) gauge theory on a 3D lattice}
\setcounter{equation}{0}

To be explicit, let us formulate the gauge model on a 
3D cubic lattice. Then the system
resembles to the lattice gauge theory introduced by Wilson\cite{wilson},
and known ample techniques are applicable to study it.
We specify each site by the site-index $x$ and use $\mu = 1,2,3$
as the direction index. We use $\mu$ also as the unit vector in the
$\mu$-th direction. We set the lattice spacing $a = 1$ for simplicity. 
For each $x$ we put a Z(2) spin variable, 
\begin{eqnarray}
S_x = \pm 1,
\end{eqnarray}
as $S_i$ with $i \leftrightarrow x$, 
and for each link $(x\mu) \equiv (x,x + \mu)$, i.e., 
for nearest-neighbor (NN) pair of sites,
we  put another Z(2) variable,
\begin{eqnarray}
J_{x\mu} = \pm 1,
\end{eqnarray}
as $J_{ij}$ with $x \leftrightarrow j, x+\mu \leftrightarrow i$. 
Below we consider the symmetric connections, 
i.e., $J_{ij} = 
J_{ji}$, so $J_{x\mu}$ describes also the conductivity of signals
propagating in the opposite direction from $x+\mu$ to $x$.
$J_{x\mu}$ is the path-dependent
phase factor of gauge theory as discussed in the previous section.
The local Z(2) gauge transformation is given by 
\begin{eqnarray}
S_x &\rightarrow& S_{x}^{'} \equiv V_x S_x, \nonumber\\
 J_{x\mu} &\rightarrow& J_{x\mu}^{'} \equiv V_{x+\mu} 
 J_{x\mu} V_{x},\nonumber\\
 V_x &=& \pm 1.
\label{z2gaugetrsf}
\end{eqnarray}

As the energy  $E$ of the system, we propose
\begin{eqnarray}
E &=& -c_1 \sum_x \sum_{\mu} S_{x+\mu} J_{x\mu} S_x\nonumber\\
&& -c_2 \sum_{x} \sum_{\mu > \nu} J_{x\mu} J_{x+\mu,\nu} 
J_{x+\nu ,\mu} J_{x\nu} \nonumber \\
&& -c_3 \sum_x \sum_{\mu} \sum_{\nu(\neq \mu)}
\big( S_x J_{x\nu} J_{x+\nu,\mu} 
J_{x+\mu,\nu} S_{x+\mu} \nonumber\\
&& +S_x J_{x-\nu,\nu} J_{x-\nu,\mu} 
J_{x-\nu+\mu,\nu} S_{x+\mu}\big). 
\label{eq:energy}
\end{eqnarray}
Each term in (\ref{eq:energy}) is depicted in Fig.\ref{fig:energy}.
The term $c_1$, which corresponds to the energy of the Hopfield model,
describes the direct transfer of signals from $x$ to $x+\mu$.
The term $c_2$ describes the self-energy of transfer of signal
through the contour $(x \rightarrow  x+\mu \rightarrow x+\mu +\nu 
\rightarrow x+\nu \rightarrow x)$ and the contour with the opposite 
direction as a reverberating circuit (\ref{reverberating}). 
The term $c_3$ describes indirect transfers of signal from $x$ to 
$x+\mu$ via the bypath, $(x \rightarrow  x+\nu \rightarrow x+\nu +\mu 
\rightarrow x$) as explained in(\ref{indirect}).
Each term of $E$ is gauge invariant, so  
\begin{eqnarray}
E(\{S_{x}^{'}\}, \{J_{x\mu}^{'}\}) &=& E(\{S_x\},\{J_{x\mu}\}).
\label{z2gaugeinv}
\end{eqnarray}


Appearance of the reverberating $c_2$ and indirect $c_3$ terms 
is also supported by the renormalization 
theory of critical phenomena. After a renormalization-group 
transformation to coarse-grain the system in the space and/or time
scales by integrating out the short-wave-length/high-frequency
modes of variables, 
every term in the energy of the resulting effective theory acquires 
corrections due to the integrated variables. 
These renormalization-group transformations also generate 
terms that are not contained in the original energy. 
In our case, if one starts with only the $c_1$-term, 
the $c_2$ and $c_3$ terms, among other terms, are
certainly generated as these effective interaction terms. 
We note that these effective terms are necessarily
Z(2) gauge-invariant reflecting
the gauge symmetry of the $c_1$ term.
This supports our postulate that the generalized model should respect
the local Z(2) gauge symmetry that has been revealed in  
the Hopfield model.

As stated in Sect.I, the present system is similar to the Z(2) 
lattice gauge theory\cite{wilson}. 
There, $S_x$ is interpreted as a Higgs 
(matter) field and $J_{x\mu}$ is the exponentiated gauge field,
$J_{x\mu} = \exp(i A_{x\mu})$. Its standard energy (action) is given by
the first two ($c_1$ and $c_2$) terms; the $c_1$ term represents
the kinetic energy of  Higgs particles interacting with the gauge field
$A_{x\mu}$, while the  $c_2$ term represents the energy of the
gauge field. Actually, in the electromagnetic
$U(1)$ gauge theory, this $c_2$ term represents the energy density
$\vec{E}^2 + \vec{B}^2$ of the electromagnetic field\cite{wilson}.
The values of these coefficients
$c_i$ are related in principle with the energy of the original 
model that is defined at the microscopic level. 
This microscopic model is, however, not easy to specify, 
so we regard these $c_i$ as effective 
parameters in a phenomenological point of view 
or in the sense of renormalization-group in the following
discussion. This set of parameters may differ individually, 
characterizing each neural network, i.e., each brain.

\subsection{Time evolution}

Let us consider the dynamics of 
$S_x(t)$ and
$J_{x\mu}(t)$.
We  postulate that the energy $E$ basically decreases as 
the time increases, which sounds natural since our brain is
a dissipative system (as long as  $E$ is regarded as the 
physical energy). This rule of energy decrease may fail
 with some rates due to misfunctions of signal processing
caused by noises, etc.,
and may be controlled by a parameter which we call
``temperature" $T \in (0,\infty)$; For higher(lower) $T$, 
failures occur more(less). This $T$ should not be confused
with the physical temperature of brain, although there may be
some correlations among them.  

For a system of continuous variables, these postulates
are implemented in Langevin equations of time evolution\cite{langevin}. 
For discrete Z(2) variables like $S_x, J_{x\mu}$, 
we propose to use the Metropolis algorithm (MA) 
\cite{MA} as the rule of time evolutions.
In fact, MA  was adopted to study the dynamics of the Hopfield
model in the name of Boltzmann machine\cite{boltzmann}.  
MA is a standard algorithm  to calculate the thermal averages 
$\langle O \rangle$
over Boltzmann distribution with the energy $E$,
\begin{eqnarray}
\langle O(S,J) \rangle &=& \frac{1}{Z} \sum_{S, J}\ 
O(S,J) \exp(-\beta E(S,J)),
\ \beta \equiv \frac{1}{T},\nonumber \\
Z &=& \sum_{S,J} \exp(-\beta E(S,J)).
\label{thermalaverage}
\end{eqnarray}
By starting  with a certain initial state $\{S_x(0),$$ J_{x\mu}(0)\}$, 
MA determines 
$\{S_x(\ell+1),$$ J_{x\mu}(\ell+1)\}$ from $\{S_x(\ell),$$ 
J_{x\mu}(\ell)\}$ by a certain probabilistic rule, and 
generates a Markov(stochastic) process,
\begin{eqnarray}
&&\{S_x(0), J_{x\mu}(0)\},
\{S_x(1), J_{x\mu}(1)\}, \{S_x(2), J_{x\mu}(2)\}, 
\cdots\nonumber\\
\label{markov}
\end{eqnarray}
In Fig.\ref{fig:MA}, we present  a flow chart of the MA for a system of  
Z(2) variables $S$.
Then it is assured\cite{MA} that the following relation holds;
\begin{eqnarray}
\langle O(S,J) \rangle  &=& \lim_{M \rightarrow \infty }
\frac{1}{M} \sum_{\ell=1}^{M} O(\{S_x(\ell), J_{x\mu}(\ell)\}).
\label{markov2}
\end{eqnarray}
We regard  the Markov process (\ref{markov}) just
as the time evolutions of
$S_x(t)$ and $J_{x\mu}(t)$
from $t$ to $t+\epsilon$ ($\epsilon$ is a certain time interval) 
with the relation $t = \epsilon\ell$.
Then the relation (\ref{markov2}) means that our time evolution
converges at large time to the equilibrium distribution 
given by (\ref{thermalaverage}). 
The time interval $\epsilon$ 
is related with the MA parameters $\alpha_S, \alpha_J$
(written as $\alpha$ in Fig.\ref{fig:MA}), 
which control the rates of
changes from $S_x(\ell)$ to $S_x(\ell+1)$ and 
 from $J_{x\mu}(\ell)$ to $J_{x\mu}(\ell+1)$, respectively;
$\alpha \rightarrow 1(0)$ implies slow(fast) changes.

As Fig.\ref{fig:MA} shows, $S_x(t+\epsilon)$ and $J_{x\mu}(t+\epsilon)$ 
are determined 
by the energy difference like $\Delta E = E(-S_x(t))-E(S_x(t)) \propto
\partial E/\partial S_x$ and 
$\Delta E = E(-J_{x\mu}(t))-E(J_{x\mu}(t)) \propto
\partial E/\partial J_{x\mu}$. These partial derivatives\cite{difference} 
are given by
\begin{eqnarray}
\frac{\partial E}{\partial S_x} &=& -c_1 \sum_{\mu} 
\big( S_{x+\mu}J_{x\mu} +S_{x-\mu}J_{x-\mu,\mu}\big) \nonumber\\
&-& c_3 \sum_{\mu} \sum_{\nu(\neq \mu)}
\big(  J_{x\nu} J_{x+\nu,\mu} J_{x+\mu,\nu} S_{x+\mu}\nonumber\\
&+&  J_{x-\nu,\nu} J_{x-\nu,\mu} J_{x-\nu+\mu,\nu}S_{x+\mu}\nonumber\\
&+&S_{x-\mu}  J_{x-\mu,\nu} J_{x+\nu-\mu,\mu} J_{x\nu} \nonumber\\
&+&S_{x-\mu}J_{x-\nu-\mu,\nu} J_{x-\nu-\mu,\mu} J_{x-\nu,\nu}
\big),\nonumber\\
\frac{\partial E}{\partial J_{x\mu}} &=&-c_1
 S_{x+\mu}S_x\nonumber\\ 
 &-& 
 c_2 \sum_{\nu(\neq \mu)}\big( J_{x+\mu,\nu}J_{x+\nu,\mu}J_{x\nu} 
+J_{x-\nu+\mu,\nu}J_{x\mu}J_{x-\nu,\nu}\big)\nonumber\\
&-&c_3\sum_{\nu(\neq \mu)}\big( S_{x-\nu}J_{x-\nu,\nu}J_{x+\mu-\nu,\nu}
S_{x+\mu-\nu}\nonumber\\
&+&S_{x+\nu}J_{x\nu}J_{x+\mu,\nu}S_{x+\mu+\nu}
+S_xJ_{x+\mu,\nu}J_{x+\nu,\mu}S_{x+\nu}\nonumber\\
&+&S_xJ_{x+\mu-\nu,\nu}J_{x-\nu,\mu}S_{x-\nu}
+S_{x+\nu+\mu}J_{x+\nu,\mu}J_{x\nu}S_{x+\mu}\nonumber\\
&+&S_{x-\nu+\mu}J_{x-\nu,\mu}J_{x-\nu,\nu}S_{x+\mu}
\big).
\label{langevin}
\end{eqnarray}
From these expressions, we observe a couple of 
 merits of the gauge-invariant energy;

(i) The time evolution of $S_x$ is affected only 
through the terms containing $J_{x\mu}$
and $J_{x-\mu,\mu}$ that have contacts with $S_x$ by inspecting
$\partial E/\partial S_x$
Likewise, the evolution of $J_{x\mu}$
is driven by the terms that have contact points
$x$ and $x+\mu$ as one inspects $\partial E/\partial J_{x\mu}$. 
The gauge symmetry assures us
that the time evolutions take place through
local interactions. As stated in Sect.II,
this is a welcome property. 

(ii) The first term in the 
right-hand-side of the second equation of (\ref{langevin})
expresses just the Hebbian rule (\ref{hebb})
to learn the present pattern $S_x$ for $c_1 > 0$.
The remaining terms in the right-hand-side describe the
indirect effects by neighboring $S_x$ and $J_{x\mu}$,
which generalize the Hebbian rule.


The typical time scales of variations
in $S_x(t)$ and $J_{x\mu}(t)$ may be different in general.
Actually, it is widely accepted that, in the actual
processes occurring in human brain, the changes in $J_{ij}$
is much slower than those of $S_i$. One may take it into 
account by controlling the characteristic time scales of 
these two sets of variables by assigning different values
for the parameters as $\alpha_S < \alpha_J$.
We come back to this point in Sect.V.

\section{Phase Structure}
\setcounter{equation}{0}

In this section, we study the phase structure of the system
with the energy (\ref{eq:energy}) at finite $T$ 
by using MA (\ref{markov}). In Sect.IVA, 
we formulate the MFT and present the resulting phase diagram 
together with the results of MC simulations. 
Some special cases of the model are also discussed.
In Sect.IVB, we explain the details of MC simulations with MA.
In Sect.IVC, we study the column-layer structure for $c_2 < 0$.

\subsection{Mean field theory and the phase diagrams}

The MFT may be formulated as a variational method\cite{feynman}
for the Helmholtz free energy $F$;
\begin{eqnarray}
Z &=& \prod_{x} \sum_{S_x=\pm 1} \prod_{x,\mu} \sum_{J_{x\mu}=\pm 1} 
\exp(-\beta E) \equiv \exp(-\beta F).
\label{partitionfunction}
\end{eqnarray}
Actually, for a trial energy $E_0$ there holds the following relations;
\begin{eqnarray}
Z_0 &=& \prod_{x} \sum_{S_x=\pm 1} \prod_{x,\mu} \sum_{J_{x\mu}=\pm 1} 
\exp(-\beta E_0) \equiv \exp(-\beta F_0),\nonumber\\
\langle O \rangle_0 &\equiv& Z_0^{-1} 
\prod_{x} \sum_{S_x=\pm 1} \prod_{x,\mu} \sum_{J_{x\mu}=\pm 1} O 
\exp(-\beta E_0), \nonumber\\ 
F &\le& F_v \equiv F_0 + \langle E -E_0 \rangle_0.
\end{eqnarray}
From this Jensen-Peierls inequality $F \leq F_v$, 
we adjust the variational parameters
contained in $E_0$ optimally so that $F_v$ is minimized.

For $E_0$ of the present system,
we assume the translational invariance
and consider the following sum of single-site and single-link energies;
\begin{eqnarray}
E_0 = - W \sum_{x}\sum_\mu J_{x\mu} - h \sum_x S_x,
\label{variationalenergy}
\end{eqnarray}
where $W$ and $h$ are real variational parameters.
Then we calculate the variational free energy per site, 
$f_v \equiv F_v/N$, where $N$ is the total number of lattice sites 
(we present the formulae for $d$-dimensional lattice) as 
\begin{eqnarray}
f_v &=& -\frac{d}{\beta} \ln( 2\cosh \beta W) -\frac{1}{\beta} 
\ln( 2\cosh \beta h) -c_1 dm^2 M  \nonumber \\
&& -c_2 \frac{d(d-1)}{2} M^4 -4c_3 \frac{d(d-1)}{2} 
m^2 M^3 + dWM + hm,\nonumber\\
m &\equiv& \langle S_x 	\rangle_0 =  \tanh \beta h, \nonumber\\
M&\equiv& \langle V_{x\mu} 	\rangle_0 = \tanh \beta W.
\label{fv}
\end{eqnarray}
The stationary conditions $\partial f_v/\partial W =
\partial f_v/\partial h = 0$ read
\begin{eqnarray}
W &=& c_1 m^2 +2c_2 (d-1)M^3 +6 c_3 (d-1)m^2 M^2, \nonumber\\
h &=& 2dc_1 mM +4c_3 d(d-1)mM^3, 
\end{eqnarray}
which give rise to the equations for $m$ and $M$;
\begin{eqnarray}
m &=& \tanh\big[2\beta d c_1 mM + 4\beta c_3 d(d-1)mM^3\big], 
\nonumber\\
M &=&   \tanh\big[\beta c_1 m^2 +2 c_2 \beta (d-1)M^3 \nonumber\\
&&+6 \beta c_3 (d-1)m^2 M^2\big].
\label{mft}
\end{eqnarray}
By assuming suitable scaling behavior of parameters $\beta c_i$
at large $d$, the result of MFT is known to become 
exact for $d \rightarrow \infty$\cite{drouffe}.

The MFT equations (\ref{fv}-\ref{mft}) generate the three phases 
characterized as follows;\\

\begin{center}
 \begin{tabular}{|c|c|c|c|c|} 
\hline
   phase    & $\langle J_{x\mu} \rangle_0 $ & $\langle S_x \rangle_0 $  
   &  ability & Hopfield
\\ \hline
Higgs       & $\neq 0$  & $\neq 0$  & learn and recall & ferromagnetic  
\\ \hline
Coulomb     & $ \neq 0$  & $0$  & learn & 
paramagnetic  \\ \hline
Confinement & $0$   & $0$     &  N.A.& N.A.   
\\
\hline
\end{tabular}

\begin{eqnarray}
\label{tab:phases}
\end{eqnarray}
\end{center}
In the first column of (\ref{tab:phases}),  
the name of each phase is given, which
is used in particle physics. The second and third columns 
show the order parameters of MFT, $\langle J_{x\mu}  
\rangle_0\ ( = M),\ \langle S_{x} \rangle_0\ (= m)$\cite{elitzur}.
The fourth column shows the properties of each phase.
The condition $M \neq 0$ is a necessary condition
so that a phase  has the ability to learn a pattern 
$S_x = \xi^\alpha_i$ by storing it to $J_{x\mu}$. 
There the fluctuations of $J_{x\mu}$ should be small
so that $J_{x\mu}$  generate a nontrivial minimum of $E$ 
at  $S_x = \xi^\alpha_i$. ($M=0$ means the fluctuations are 
too large.)
The condition $m \neq 0$ is a necessary condition to recall 
the pattern $S_x = \xi^\alpha_i$. The fluctuations of
$S_x$ around $S_x = \xi^\alpha_i$ should be small 
when recalling is successful. We note that the  
fourth phase with  $M = 0$ and $m \neq 0$ is missing.
This sounds natural since learning should be a necessary 
condition for recalling, hence partly supporting the gauge 
symmetry of neural network.
The fifth column indicates the corresponding phases in 
the Hopfield model.

In Fig.\ref{fig:phasediagram} we plot the phase diagrams 
obtained from (\ref{fv}-\ref{mft}) for various values of $c_3$. 
(The case of $c_3 = 0$ has been studied in Ref.\cite{matsui}.)
The results of MC simulation in the next subsection are also 
presented by filled circles. 
The phase boundary of MFT between the Higgs phase and the Coulomb
phase is second order, while other two boundaries, 
Higgs-confinement and confinement-Coulomb, are first order.
In Fig.\ref{fig:fv} we present typical behaviors of $f_v$ for 
(a): second-order transition; (b) first-order transition.
Across a second-order transition, $M$ and $m$ vary continuously,
while across a first-order transition, $M$ and/or $m$ change 
discontinuously with finite jumps $\Delta M$ and/or
$\Delta m$. For a Higgs-confinement transition,
$\Delta M \neq 0$ and $\Delta m \neq 0$, and
 for a confinement-Coulomb transition,
 $\Delta M \neq 0$ and  $\Delta m = 0$ since $m = 0$ in both phases.

The locations of phase boundaries of MFT 
agree globally with those of MC simulation in Sect.IVB. 
However, the results of MFT are not sufficient 
in the following two points;

(i)  The MC simulation shows that 
the confinement-Coulomb transitions are second-order contrary to the MFT.
This point may be explained as $d=3$ is not large enough
for MFT. The MC simulation for $d =4$ \cite{creutz} for $c_3 = 0$
reports first-order confinement-Coulomb transitions as the MFT does.

(ii) The MC simulation shows that the Higgs-confinement boundary 
terminates at a certain end point.
Along this phase boundary, the jumps $\Delta M, \Delta m$ 
decrease as $c_2$ decreases 
and disappear at the end point at $c_2 > 0$. 
This corresponds to the complementarity studied in 
Ref.\cite{complementarity} for $c_3 = 0$, which states that
 these two phases are analytically connected through a detour.

There are the following special cases that are examined better 
by other methods;

\begin{center}
{ CASE\ I: Ising model}\ $(c_2 \rightarrow \infty)$\\ 
\end{center}
Here the $c_2$ terms forces
$J_{x\mu}$ to the so-called pure-gauge 
configuration, and the energy reduces to that of the Ising model;
\begin{eqnarray} 
J_{x\mu} &\rightarrow& V_{x+\mu}V_x,  \nonumber\\
E &\rightarrow& \ -(c_1 + 4 c_3) \sum_x \sum_{\mu} S'_{x+\mu} S'_x 
+ {\rm const.},\nonumber\\
S'_x &\equiv& V_x S_x = \pm 1. 
\label{eq:isingenergy}
\end{eqnarray}
Thus, there is a second-order Ising transition at 
$\beta(c_1 + 4 c_3) \simeq 0.22$ for $c_2 =\infty$.\\

\begin{center}
{ CASE\ II:\ Pure gauge model}\ $(c_1 = c_3 = 0)$\\ 
\end{center}
Here the energy is
\begin{eqnarray} 
E &= & -c_2 \sum_{x} \sum_{\mu > \nu} J_{x\mu} J_{x+\mu,\nu} 
J_{x+\nu ,\mu} J_{x\nu} + {\rm const}.
\label{eq:gaugeenergy}
\end{eqnarray}
This system is known \cite{creutz} to exhibit a second-order transition
at $\beta c_2 \simeq 0.76$. Actually, after the duality transformation, 
the system (\ref{eq:gaugeenergy}) is converted to the 3D 
Ising spin model.\\

\begin{center}
{ CASE\ III:\ Single-link  model}\ $(c_2 = c_3 = 0)$\\ 
\end{center}
Here the sum over $J_{x\mu}$ is possible because
the energy is decoupled to each link, and the partition function 
becomes an $S_{x}$-independent constant; 
\begin{eqnarray} 
\sum_{J_{x\mu}= \pm1} \exp(\beta c_1 S_x J_{x\mu} S_{x+\mu}) &= &
2 \cosh(\beta c_1S_x S_{x+\mu})  \nonumber\\
&=&2 \cosh(\beta c_1),
\label{eq:singlelinkenergy}
\end{eqnarray}
due to $S_x^2 = 1$. Thus the free  energy has no singularities
and no phase transitions along the $c_1$-line. This explains
why the Higgs-confinement transition line should 
truncate at the end point (before reaching the $c_1$-line) 
as the complementarity\cite{complementarity} claims.

\begin{center}
{ CASE\ IV: Self-duality curve}\ $(c_3 = 0)$\\ 
\end{center}
For $c_3 = 0$, the standard Z(2) duality transformation
can be  applied\cite{wegner}, which
maps the model at $P(\beta c_1,\beta c_2)$
onto the same model at $P'(\beta c_1',\beta c_2')$ with
\begin{eqnarray}
\beta c_1' &=& -\frac{1}{2}{\rm \ell n th}(\beta c_2), 
\
\beta c_2' = -\frac{1}{2}{\rm \ell n th}(\beta c_1). 
\label{self-duality}
\end{eqnarray}
If we {\it assume} the phase transition occurs at the point
satisfying $P' = P$, we  obtain the curve of phase transition
from (\ref{self-duality}) as 
\begin{eqnarray}
\beta c_2 &=& -\frac{1}{2}{\rm \ell n th}(\beta c_1). 
\label{self-duality-pt}
\end{eqnarray}
In the next subsection, we shall see that {\it a part } of 
the curve (\ref{self-duality-pt}) is actually the phase boundary.

\subsection{MC simulation}

We performed MC simulations for a 3D lattice of the size 
$N = L^3$ up to $L = 16$ with the periodic boundary condition.
The case of $c_1 = c_3 = 0$ has been examined
by Bahnot and Creutz.\cite{creutz}.  
We employed MA, which is illustrated in Fig.\ref{fig:MA},
with choosing the prefactors $\alpha_S = \alpha_J = 0.9$.
Typical numbers of sweeps [$M$ of (\ref{markov})] 
are $10^5$ for thermalization and $5 \times 10^4$ for 
measurements.

Among others, we measured the internal energy $U$ and 
the specific heat $C$,
\begin{eqnarray}
U &=& \langle E \rangle, \nonumber\\
C &=& \frac{dU}{dT} = \beta^2 \langle\ 
( E-\langle  E \rangle\ )^2 \rangle.
\end{eqnarray}
We judge the order of transition as follows;
If $U$  has a discontinuity at the transition point, 
it is of first order. A typical behavior of $U$ and $C$ for 
a first-order transition is given in Fig.\ref{fig:Cfirst}a and 
Fig.\ref{fig:Cfirst}b.  If $U$ is continuous and
$C$ has a peak and discontinuity, then it is of second order. 
A typical behavior of $U$ and $C$ for  a second-order transition
is given  in  Fig.\ref{fig:Csecond}a and Fig.\ref{fig:Csecond}b.

Let us comment on the gauge fixing. 
Our MC simulations  have been done without fixing the gauge.
We  have also made MC simulations in the unitary gauge,
\begin{eqnarray}
S_x = 1.
\end{eqnarray}
Although the partition function
in (\ref{partitionfunction}) and the corresponding averages are 
independent whether one fixes the gauge or not, 
variations  of the variables in Markov processes 
in the unitary gauge are too small
and the convergence of the expectation values are too slow 
to obtain meaningful results with good accuracies.  
This slow convergence shall persist even for other gauge fixings.

\subsection{Column-layer structures in $c_2 < 0$}

Usual studies of lattice gauge theory have been restricted to 
the case of $c_2 > 0$.
However, as a model of neural network, the case of 
$c_2 < 0$ is also interesting since this condition implies that 
the $c_2$-term in the energy expresses negative feed backs, 
that is, signals starting from $x$ will propagate around  
the plaquette and return to $x$ with a negative coefficient.

For $c_2 < 0$, the $c_2$-term prefers ``flux
states" of $J_{x\mu}$, i.e., 
$J_{x\nu}J_{x+\nu,\mu}J_{x+\mu,\nu}J_{x\mu} = -1$. Then it will
generally compete with the $c_1$-term which 
prefers fluxless states, 
$J_{x\nu}J_{x+\nu,\mu}J_{x+\mu,\nu}J_{x\mu} = 1$\cite{fluxless}.
Thus the system is frustrated.
The MFT with the translationally-invariant variational energy 
of (\ref{variationalenergy}) 
is inadequate for this situation, and the 
MC simulation is necessary.

In Fig.\ref{fig:phasediagram2}, 
we present the phase diagram of MC simulation
for the extended region with $c_2 < 0$ and/or $c_1 <0$. 
In the region $c_2 < 0$, there are phase boundaries that look 
like mirror images of $c_2 > 0$ of 
Fig.\ref{fig:phasediagram} except that
the Higgs-confinement boundaries extend to $c_1 \rightarrow 
\pm\infty$ instead of terminating at the ending critical points. 
The phase F in Fig.\ref{fig:phasediagram2} corresponds to the
Higgs phase B, the phase E to Coulomb phase A, 
and the region D is connected to the confinement phase C.
 For $c_3 = 0$ the phase diagram is symmetric w.r.t the 
 $c_1=0$-line, reflecting the symmetry under
$(c_1,c_3) \leftrightarrow (-c_1,-c_3)$\cite{symmetry}, i.e., 
\begin{eqnarray}
&& Z(c_1,c_2,c_3) =  Z(-c_1,c_2,-c_3).
\label{symmetry}
\end{eqnarray}
Bahnot and Creutz\cite{creutz2} performed MC simulations
for  $c_2 < 0$ at $c_3= 0$ with the related interest of spinglass
(though taking $J_{x\mu}$ as an annealed variable). 
Our result at  $c_3= 0$ is consistent 
with their result. At $c_3 \neq 0$, the symmetry w.r.t. the 
$c_1=0$-line is violated as shown in Fig.\ref{fig:phasediagram2}b 
for $\beta c_3 = 0.05$.

In the ``Higgs" phase F, we observe a ``column-layer" structure
in the spatial configurations of the following link objects;
\begin{eqnarray}
 j_{x\mu} \equiv \langle S_{x+\mu}J_{x\mu}S_x \rangle.
\end{eqnarray}
$j_{x\mu}$ is the thermal average of
a gauge invariant extension of $J_{x\mu}$ 
as appeared in the $c_1$-term of the energy.
It measures the efficiency of signal propagations between
$x$ and $x+\mu$.
In Fig.\ref{fig:layer}, we present the basic unit of a typical
periodic configurations of  $j_{x\mu}$ that forms 
a column-layer lattice in the phase F. 

This column-layer structure on the lattice is stable, 
because $j_{x\mu}$ are not snap shots  but thermal (i.e., time) 
averages in long MC runs. It is to be spontaneously generated
as a result of dynamics, i.e., a result of  self-organization 
in the present model. In short, its origin is the frustration 
(competition) between the $c_1$ and $c_2$ terms in the energy.

There are several possible configurations of $j_{x\mu}$
which differ from that of Fig.\ref{fig:layer} in the 
direction of 1D alignments (i.e., other than in the 1-direction) 
and the orientation of the planes of layers (other than
in the 1-2 plane).  These options are generated by starting with 
different initial configurations (and/or different random
numbers). In this sense, their generations are spontaneous,
but they are stable once they are formed as we explained.

Since a large $j_{x\mu}$ means that signals
 (potential) propagate between $S_x$ and $S_{x+\mu}$ frequently
 and coherently, each column in the column-layer structure 
 is a 1D path along
which signals propagate dominantly. 
It is interesting to study possible relevance of such a
 structure to the actual self-organized structures 
 observed in the human brains like ocular dominance columns.
 \cite{selforganization,malsburg,tanaka,linsker} (See the future
 problems in Sect.VI.)
We also note that the present 
structure of weakly-coupled planes may be viewed as a kind of 
multilayer neural network.


\section{Learning and Associative Memory}
\setcounter{equation}{0}

In the previous section, we studied the phase structure
of the model, which reflects the static properties of the model.
In this section, we simulate the processes of learning a pattern
of $S_x$ and recalling it. The results reflect the dynamical
properties of the model.

We set up the simulations in the following two steps in time;\\

{(i) Learning during $0 < t < t_1$:}\\

\noindent
We first prepare a pattern
$S_x = \xi_x$ to memorize and start with it, $  S_x(0)=\xi_x $.
During this learning time, we freeze $S_x(t)$ as  
\begin{eqnarray}
S_x(t) &=& S_x(0) \ {\rm for }\ 0 < t < t_1,
\end{eqnarray} 
by hand to let the system learn it. 
This may correspond to apply very strong stimuli to the brain
like forcing it to watch the pattern with concentration. 
On the other hand,
we allow $J_{x\mu}(t)$ vary according to MA to adjust themselves to a 
configuration suitable for $S_x(0)$.
In terms of the prechoice parameters $\alpha_S, \alpha_J$ of MA
(where $1-\alpha$ is a parameter
to control the rate of time variation),
we set $\alpha_S = 1$ and $\alpha_J < 1$ for $0 < t < t_1$.
Below we fix $\alpha_J = 0.9$ for $0 < t < t_1$.
The time $t$ is defined throughout
the simulations ($0 < t$) as $t  \equiv (1-\alpha_J(0 < t < t_1))
\times $ number of iterations (sweeps), i.e.,
$t  \equiv 0.1 \times $ number of iterations.\\

{(ii) Recalling during $t_1 <  t$:}\\

\noindent
  At $t = t_1$ we change $S_x$  discontinuously
from $S_x(0)$ to a pattern $S_x(t_1)$    that is obtained 
by adding a certain amount of random noise upon $S_x(0)$.
For definiteness, we set $S_x(t_1) = -S_x(0)$ for the 10\% 
of sites $x$ that are randomly chosen throughout the entire 
lattice (except for Fig.\ref{fig:Osnoise} discussed later 
where we consider the cases of more amounts of noises). 
This forced change of $S_x(t)$ at $t= t_1$ 
simulates relaxing the brain from the concentration upon 
$S_x(0)$ and letting it forget  $S_x(0)$ partly.  
Then we let $S_x(t)$ vary according to MA with $\alpha_S < 1$.
Below we fix $\alpha_S = 0.9$ for $t_1 < t$.
On the other hand,  $J_{x\mu}(t)$ basically vary according to MA
with $\alpha_J < 1$ smoothly starting from $J_{x\mu}(t_1)$.
However, with the reason explained below, 
we  shall be  also interested in the special limit
$\alpha_J \rightarrow 1$ in which  the time 
variations of $J_{x\mu}(t)$ after $t_1$ freeze;
\begin{eqnarray}
\alpha_J \rightarrow 1:\ 
J_{x\mu}(t) &=& J_{x\mu}(t_1) \  {\rm for }\  t_1 < t.
\label{freezedj}
\end{eqnarray} 

To judge the performance of each process of 
learning a pattern and recalling it in a quantitative manner, 
we use the following overlaps;
\begin{eqnarray}
O_S(t) & \equiv& \frac{1}{N} \sum_x S_x(0)S_x(t),\nonumber\\
O_J(t) & \equiv& \frac{1}{3N}\sum_x \sum_\mu J_{x\mu}(0)J_{x\mu}(t),
\end{eqnarray}
which are gauge-invariant under (\ref{z2gaugetrsf}).
If the recalling is successful, we expect $S_x(t) \simeq S_x(0)$
at sufficiently large $t$, so $O_S \simeq 1$.
(Note that $O_S(t_1) = 0.8$ for the 10\% change of $S_x$ at $t= t_1$.)
On the other hand, $O_J(t_1)$ measures the rate how much the synaptic 
connections change by learning during $0 < t <t_1$.
In Fig.\ref{fig:typicalOs}, we illustrate $O_S$ and $O_J$ versus $t$ 
in two typical processes for the case (\ref{freezedj}).
Fig.\ref{fig:typicalOs}a is  a process that 
succeeds in recalling, 
and Fig.\ref{fig:typicalOs}b is one that fails in recalling.

Let us first study the dependence of the overlaps
upon the choice of $\alpha_J$ for $t_1 < t$. 
In Fig.\ref{fig:forget} 
we plot $O_{S}(t)$ and $O_{J}(t)$ for three cases, $\alpha_J = 1.0, 
0.9, 0.99$ for $t_1 < t$. (Note that we fixed
$\alpha_J = 0.9$ for $0 < t < t_1$.) As the initial condition,
we choose $S_x(0)$ and $J_{x\mu}(0)$ randomly. 
We select three points in the parameter space of $c_i$.
Fig.\ref{fig:forget}a is the result for 
$c_1 = 1, c_2 = c_3 =0$ which belongs to
the confinement phase.  For $\alpha_J = 1$
($J_{x\mu}(t) = J_{x\mu}(t_1)$),
$O_S$ approaches to the constants near 0.95;
the system (almost) succeeds to recall $S_x(0)$.
For  $\alpha_J = 0.99$, $O_S$ once increases 
and then decreases. It describe a process
of recalling $S_x(0)$ partly and then lose the memory gradually,
that is, the phenomenon of a partial memory loss(deficit).
For $\alpha_J = 0.9$, $O_S$ decreases monotonically.
These latter two curves exhibit
a typical phenomenon of a dynamical system
with coupled variables. A fixed point obtained by fixing some variables
may become unstable when all are coupled.
As expected, $O_S$ decreases more as $J_{x\mu}$ 
change more rapidly (i.e., $\alpha_J$ decreases more). 
Fig.\ref{fig:forget}b is for the point 
$c_1 = 3.0, c_2= 1.0, c_3 = 0$, and Fig.\ref{fig:forget}c 
is for the point $c_1 = 3.0, c_2= -1.0, c_3 = 0$, both of 
which are in the Higgs phase. 
In contrast to Fig.\ref{fig:forget}a, $O_S$ in these points 
have smaller dependence on $\alpha_J$, so the 
effects of memory loss 
are  smaller than in the confinement phase. 
Although the time-variations of $J_{x\mu}(t)$ for
$t_1 < t$ ($\alpha_J \neq 1$)
reduce the performance of recalling more (in the deconfinement
phase) or less (in the Higgs phase), 
we stress that these phenomena of partial memory loss are not a flaw 
of the present model, but a welcome feature that the realistic
model of human brain should possess. 

Since the case of $\alpha_J = 1$
for $t_1 < t$ gives the stable and highest values of $O_S$
at large $t$, we present the results for  
$\alpha_J(t_1 < t) = 1$ 
in Fig.\ref{fig:Osinitial}-\ref{fig:Oscontour}\cite{ajless1}
below as typical results for $\alpha_J \simeq 1$.
This choice $\alpha_J(t_1 < t) = 1 $ for presentations
sounds also reasonable because 
the actual variations of synaptic connections, $J_{x\mu}(t)$,
 in the human brain are
 much slower than those of $S_x(t)$ as explained 
in Sect.IIIB, i.e., $\alpha_J \simeq 1$.
(In this viewpoint, our choice $\alpha_J = 0.9$ for $0 < t < t_1$ 
corresponds to accelerate the learning processes significantly.
If we take larger $\alpha_J \simeq 1$ for $0 < t < t_1$, then we need
to take $t_1$ larger in order to obtain the same values of $O_S$.)

Next we study the dependence on the initial conditions.
In Fig.\ref{fig:Osinitial} we present the contours of 
$O_S$ at large $t$ for four choices (a-d) of the initial states 
$S_x(0), J_{x\mu}(0)$.
The results of first three choices (a-c) look similar each other,
while the fourth case (d) $S_x(0)=1,  J_{x\mu}(0)=1$ is different;
the region of higher
$O_S$ for $c_2 > 0$ extends to the phase boundary $(c_1 \simeq 0.4)$ 
of the Higgs phase and the Coulomb phase.
This is natural since the configuration  $S_x =J_{x\mu}=1$
is the minimum of the free energy in the Higgs phase at $c_2 >0$. 
Thus the system has learned
this pattern $S_x(0)= 1$ from the beginning.
Below we present the result for the random initial condition,
the choice (a), i.e., $S_x(0)$ and $J_{x\mu}(0)$  are chosen 
randomly from $\pm1$.
 
To study the dependence on the learning time $t_1$,
we present in Fig.\ref{fig:Ost1} two typical processes with 
different $t_1$. 
In Fig.\ref{fig:Ost1}a, $t_1 = 1$ and 
$O_S$ approaches to $O_S=0.6$, so it fails to recall $S_x(0)$.
On the other hand, in Fig.\ref{fig:Ost1}b, $t_1 = 5$ and 
$O_S$ approaches to $O_S=0.92$, so we judge
it succeeds  (but not completely) to recall $S_x(0)$.
A reason is clearly drawn from the behavior of $O_J$. 
In Fig.\ref{fig:Ost1}b, $O_J$ 
almost converges to a fixed value at $t_1$, that is 
$J_{x\mu}$ converge
to the suitable configuration,
while, in Fig.\ref{fig:Ost1}a, $O_J$ is in a way to converge. 
Thus, a necessary condition to learn successfully is to keep 
$t_1$ sufficiently long (longer than the relaxation time of $J_{x\mu}$) 
so that $J_{x\mu}$ can converge to the configuration that
makes $S_x = S_x(0)$ a minimum of the energy. 
For definiteness we present the results for $t_1 = 5$ below.

Let us next  study the relative importance of each term $c_1,c_2,c_3$
of the energy in learning and recalling. 
In Fig.\ref{fig:Osc123}, 
$O_S$ after sufficiently large time is plotted for three cases
 where only one $c_i$ of $c_1,c_2,c_3$
 is nonvanishing and the other two are zero. 
In the case of $c_1$ alone, as $c_1$ increases, $O_S$ approaches unity. 
This is expected
since  the $c_1$-term describes the direct transfers of signals. 
In the case of $c_3$ alone, as $c_3$ increases,  
$O_S$ increases but saturates around $O_S \simeq
0.65$. This indicates that the indirect signal transfers by
the $c_3$-term is not sufficient by itself to recall the original 
pattern, as anticipated. In the case of $c_2$ alone, 
$O_S\approx 0$. This is natural because the $c_2$-term,  which
contains only $J_{x\mu}$ and no $S_x$, expresses signal-transfers 
starting from a neuron and ending at the same neuron through
a loop along a plaquette, but describes no signal-transfers 
to the NN neurons.

Let us see the roles of $c_2,c_3$-terms in details. Since we have
observed that the $c_1$-term plays the central role in learning and 
recalling, we simulate the processes with $c_1 + c_2$ and $c_1 + c_3$.
In Fig.\ref{fig:Osnoise} we plot $O_S$ after large time for several 
$S_x(t_1)$'s prepared by adding certain amounts of noises to $S_x(0)$
from 10\% up to 40\%. 
It shows that adding certain amounts of 
$c_2$ or $c_3$ upon $c_1$ improves the rate of recalling, 
i.e., to achieve larger $O_S$.
It is interesting to note that there is a preferred region for $c_2$,
$\beta c_2 \in (0.2, 0.6),$ for which $O_S$ is over 95\%.
This reminds us the phenomenon that applying a certain 
but not too much amount
of magnetic field improves our brain activities.

 Let us study the relation between the results of present section and 
the phase structure of the previous section.
In Fig.\ref{fig:Oscontour}, we superpose contour plots of
$O_S$ on the phase diagram Fig.\ref{fig:phasediagram}b.
We observe that being in the 
Higgs phase is not sufficient to achieve good rates of successful
learning and recalling. This is clear especially in 
the region with large $c_2$, which is consistent with the case of
$c_2$ alone in Fig.\ref{fig:Osc123}. 
Also, in the region near $\beta c_1=0.5, \beta c_2=0.4$
which is a vague border between the  confinement and the Higgs phases, 
learning is possible. 
 So this region may have something
 to do with our experience that a tiny amount of stimulation
 helps us to recall certain patterns; a coexisting phenomenon
 of recalling and nonrecalling. 
 Apart from these regions, there are certainly the correlations
 among the results of this section and the phase diagram of Sect.III. 

From these results,  one may list up the conditions to succeed in
learning a pattern of $S_x$ and recalling it as follows;\\

\noindent
- The learning time $t_1$ should be 
larger than the relaxation time of $J_{x\mu}$.

\noindent 
- The temperature $T$ should be low.

\noindent
- The self-interaction term $\beta c_2$ should be within 
a certain range ($0.0 < \beta c_2 < 0.7$ for $\beta c_1=1.0$).

\noindent
- The indirect  $c_3$-term should be of the same signature 
as the direct $c_1$-term
to accelerate signal transfers caused by the $c_1$-term.

\section{Conclusions and Future Problems}
\setcounter{equation}{0}

By converting the strength $J_{ij}$ of the synaptic connection 
to a Z(2) gauge variable (exponentiated gauge connection) and 
imposing the gauge symmetry on the energy 
$E(\{S_i\},\{J_{ij}\})$,
we have proposed an explicit model of neural network of learning.
The Z(2) gauge symmetry is inherited from the Hopfield model. 
Study of the phase structure and the simulations of learning and 
recalling revealed several interesting features of the model. 

\noindent
- In the confinement phase, both learning and recalling  
are disabled, which may corresponds to certain symptoms like
Alzheimer's disease [See (\ref{tab:phases})].
The complementarity characterized by the end point of 
Higgs-confinement phase-transition curve may offer us some methods to
retrieve the network from the confinement phase to the Higgs phase
in a continuous and practical manner.

\noindent
- The column-layer structure of $\langle S_{x+\mu}J_{x\mu}S_x\rangle$
discussed in Sect.IVC implies that there are particular 1D paths 
(columns) in each layer along which the signals (electric potential) 
propagate dominantly. This structure seems to exhibit the potentiality
of the present model to evolve the self-organized column structures
of the active neurons, which are observed in the human 
brains\cite{selforganization,malsburg,tanaka,linsker}. 

\noindent
- Due to the mutual interactions between
$S_x$ and $J_{x\mu}$, the phenomena of memory loss are observed
as in Fig.\ref{fig:forget}. 

\noindent
- The time evolution of $J_{x\mu}$ generalizes the Hebbian rule
of learning. Suitable amounts of the newly added 
$c_2,c_3$-terms improve the 
performance of learning and recalling (Fig.\ref{fig:Osnoise}).

We think that these features make the present model interesting 
as a neural network of learning and associative memory
and worth to make further investigations.

There are certainly various ways to improve the present model.
Among others, we list up the following;

(1) Actual synaptic connections are not symmetric, 
i.e., $J_{ij} \neq J_{ji}$. However, in the Hopfield model, 
due to the very form of its energy $E$ of (\ref{hopfieldenergy}), 
the antisymmetric part of $J_{ij}$, $J_{ij}-J_{ji}$, does not 
contribute to $E$. This flaw can be removed in the present framework
by introducing {\it two independent Z(2) variables},
 $J_{x\mu}$ and $J_{x+\mu, -\mu}$, on a link $(x,x+\mu)$. 
Then the $c_2$ and $c_3$ terms can reflect the antisymmetric 
part  $J_{x\mu} - J_{x+\mu, -\mu}$. 
In fact, an asymmetric model in this direction is proposed 
in Ref.\cite{matsui} (Model III) and its phase diagram is studied
in MFT.


(2) In the present model, synaptic connections are restricted only to
the NN neurons. In  human brain, each neuron receives signals
from $1000 \sim 10000 $ neurons.
These long-range connections are certainly responsible to
store many patterns and should be incorporated in a more 
realistic model.
We plan to increase the number of connections in the future study.
This means to introduce more variables $J_{ij}$ beyond
NN pairs.
Related to this point, one may increase the number of patterns to 
memorize as $\xi_x^{\alpha}$ in the Hopfield model.


(3) Natural ways to incorporate external stimuli like visual 
and acoustic ones to the present model may be (i)  
to change the boundary condition 
from the present periodic boundary condition to a fixed boundary
condition with an appropriate values of $S_x, J_{x\mu}$ 
on the surface of the 3D lattice, and/or (ii) fixing $S_x, J_{x\mu}$
in certain part of the system to certain constants.
By studying the response to each stimulation, one may address the 
question whether column structures of active neurons are 
generated in the present model. \\


\begin{center} 
{\bf Acknowledgments} 
\end{center} 
We thank Yukari Fujita and Yuki Nakano for discussion.
One of the authors (K.S.) thanks the members of 
Department of Physics, Kanazawa University for their 
hospitality delivered to him during his stay.

\begin{center} 
{\bf Note added}
\end{center} 
The gauge model of neural network has been extended
to a quantum neural network having U(1) gauge symmetry in
Y.Fujita and T.Matsui,  
Proceedings of 9th International Conference on Neural 
Information Processing, ed. by L.Wang et al. (2002)1360-1367 
(arXiv:cond-mat/0207023).



\renewcommand{\thefigure}{\arabic{figure}}
\newpage

\begin{figure}
  \begin{picture}(210,160)
    \put(50,0){\epsfxsize 160pt \epsfbox{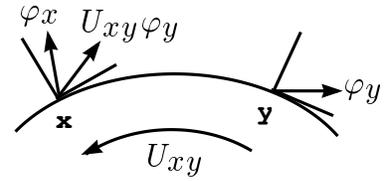}}
  \end{picture}

\vspace{-1.5cm}
  \caption{\ Function of gauge variable $U_{xy}$ (path-dependent
  phase factor).
$U_{xy}$ parallel-translates a vector $\varphi_y$ at the point $y$ 
to another point $x$ giving rise to $U_{xy}\varphi_y$. To compare
two vectors $\varphi_x$ and $\varphi_y$, one needs
to refer to the common local frame, say the frame at $x$.
So  one should 
parallel-translate $\varphi_y$ to $x$ and take the gauge
invariant scalar
product $(\varphi_x, U_{xy} \varphi_y) \equiv \varphi^\dagger_x U_{xy} 
\varphi_y$ instead of $(\varphi_x, \varphi_y) =\varphi^\dagger_x 
\varphi_y$.
 }
\label{fig:gauge}
\end{figure}

\begin{figure}
  \begin{picture}(210,160)
    \put(30,40){\epsfxsize 140pt \epsfbox{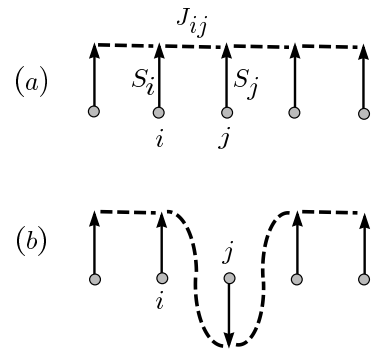}}
  \end{picture}

\vspace{-1cm}
\caption{\ 
Illustration of the local frames of $S_i$. 
Each solid line with an arrow indicates  $S_i (=\pm1)$ at 
the site $i$ (filled circle).  
Each dashed line indicates
the gauge variable $J_{ij}$ which measures the relative orientation
of local frames at nearest-neighbor points. 
(a) Localframes where all
the frames are in the same orientation.    
(b)  Local frames after the local gauge transformation at $j$,
$S'_j = -S_j$. The information of 
this change is stored(absorbed) in the neighboring
gauge variables $J_{ij}$ so that the signal received at $i$
is unchanged  $J'_{ij}S'_j = J_{ij}S_j$.
 }
\label{fig:localframe}
\end{figure}


\begin{figure}
  \begin{picture}(210,100)
    \put(0,40){\epsfxsize 230pt \epsfbox{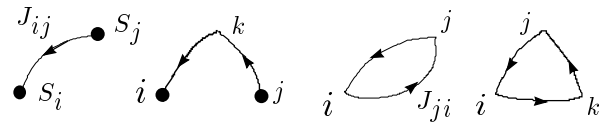}}
  \end{picture}

\vspace{-1cm}
  \caption{\ Examples of gauge-invariant terms.
  The black circles denote $S_i$ and the curves with arrows
  denote $J_{ij}$. The figures indicate $S_i J_{ij} S_j,\ 
  S_i J_{ik} J_{kj} S_j,\ J_{ij}J_{ji},\
  J_{ij}J_{jk}J_{ki}$ from
  the left, respectively. The first $SJS$ term is just the term
  of the Hopfield model. The second $SJJS$ term
  may be viewed to describe the two successive processes
  $S_k J_{kj} S_j$ and $S_i J_{ik} S_k$.
  The last two terms consist of $J$'s only, and describe
  closed circuits, which may be taken as ``reverberating circuits".  
 }
\label{fig:energy0}
\end{figure}

\newpage

\begin{figure}
  \begin{picture}(210,160)
    \put(30,0){\epsfxsize 180pt \epsfbox{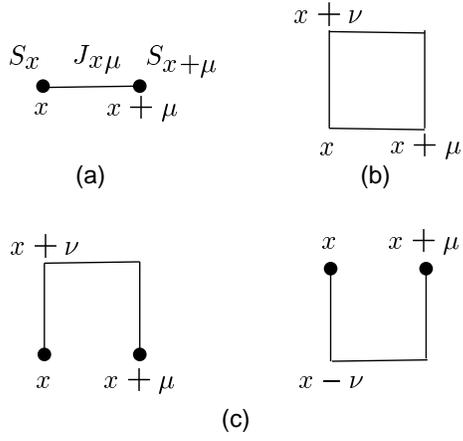}}
  \end{picture}
  \caption{Graphical representation of each term in $E$ of 
  (\ref{eq:energy}).
The black circles represent $S_x$ and the line segments represent
$J_{x\mu}$. Fig.a describes the direct interaction
between $S_x$ and $S_{x+\mu}$ through $J_{x\mu}$. 
Fig.b is taken as the smallest ``reverberating circuit".
Fig.c describes the combined effects of three successive
processes given by Fig.a.}
\label{fig:energy}
\end{figure}


\begin{figure} 
  \begin{picture}(200,230) 
    \put(40,0){\epsfxsize 170pt \epsfbox{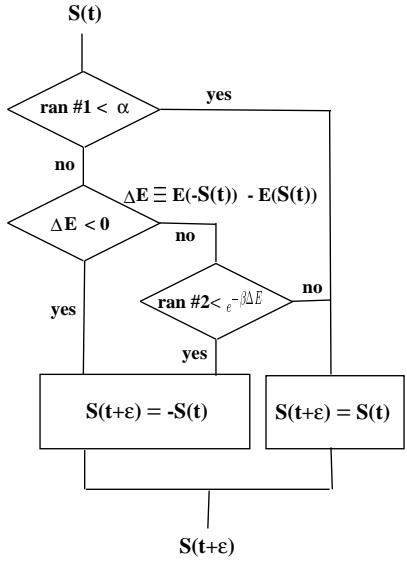}} 
  \end{picture} 
  \caption{ Flow chart of Metropolis algorithm for
  a system with Z(2) variables $S =\pm1$ to determine
  the value $S(t+\epsilon)$ starting from $S(t)$.
  ran\#1, ran\#2 are random numbers distributed uniformly
  in the interval $(0,1)$. 
  This update process is to be done
  for each  variable for every time step. 
  In the present model,  each update  at every time step  
  sweeps out all $S_x(t), J_{x\mu}(t)$
  throughout the entire lattice.
} 
\label{fig:MA}
\end{figure} 

\vspace{8cm}

\vspace{-0.7cm}
\begin{figure} 
\begin{picture}(200,140) 
    \put(0,0){\epsfxsize 200pt \epsfbox{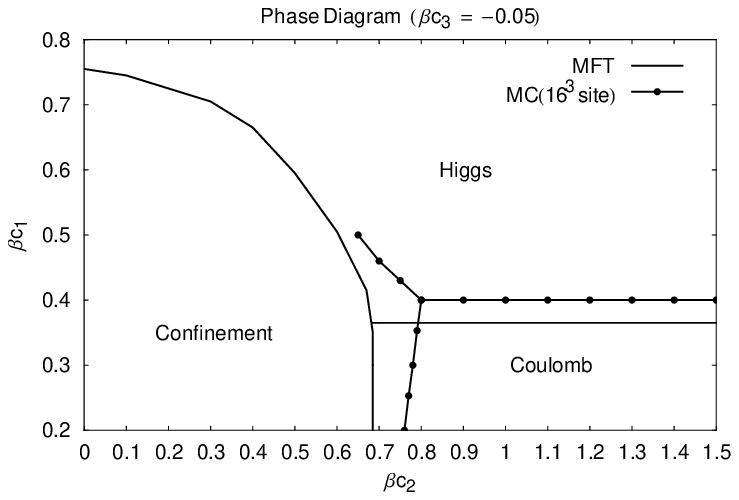}}
\end{picture} 
\begin{center}
\vspace{-0.2cm}
(a)
\end{center}

\vspace{-0.7cm}
\begin{picture}(200,140) 
    \put(0,0){\epsfxsize 200pt \epsfbox{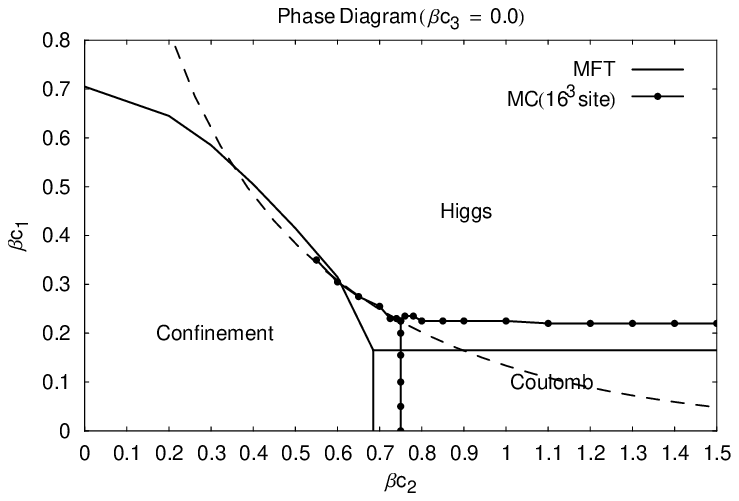}} 
  \end{picture} 
\begin{center}
\vspace{-0.2cm}
(b)
\end{center}

\vspace{-0.7cm}
\begin{picture}(200,140) 
    \put(0,0){\epsfxsize 200pt \epsfbox{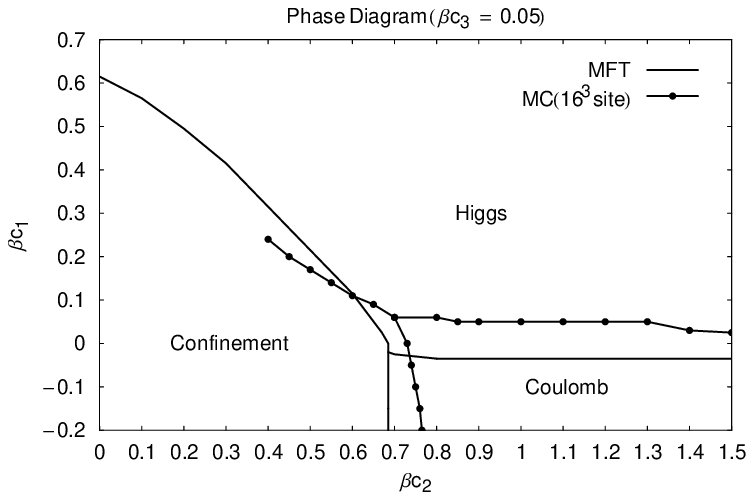}} 
  \end{picture} 
\begin{center}
\vspace{-0.2cm}
(c)
\end{center}

\vspace{-0.7cm}
\begin{picture}(200,140) 
    \put(0,0){\epsfxsize 200pt \epsfbox{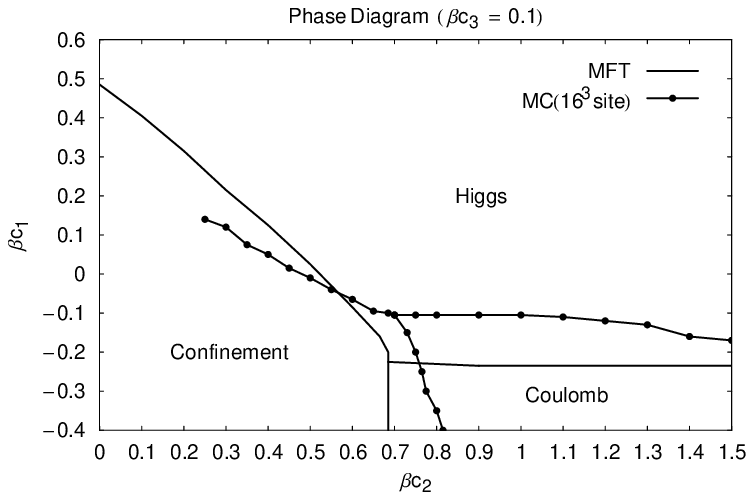}} 
  \end{picture} 
\begin{center}
\vspace{-0.2cm}
(d)
\end{center}

\caption{ 
Phase diagram by MFT and MC simulation. (a) $\beta c_3=-0.05$,
(b) $\beta c_3=0.0$, (c) $\beta c_3=0.05$, (d) $\beta c_3=0.1$.
MC simulations show that the Higgs-Coulomb transition 
and the confinement-Coulomb transition are second order,
while the Higgs-confinement transition is first order.
The MFT predicts the first-order confinement-Coulomb transition
incorrectly. Also the Higgs-confinement boundaries terminate
at certain critical points instead of extending to $c_2 = 0.$ 
The dashed curve in (b) is the  phase-transition curve
(\ref{self-duality-pt}) predicted by the duality transformation.
It almost agrees with our MC result in the period 
$ 0.55 < \beta c_2 < 0.75$,
in which the assumption $P'=P$ for phase transitions 
is verified.} 
\label{fig:phasediagram} 

\end{figure} 

\begin{figure}
  \begin{picture}(190,120)
    \put(30,0){\epsfxsize 180pt \epsfbox{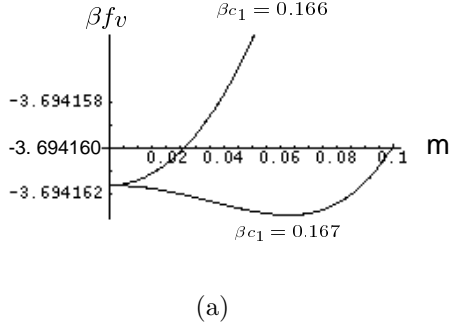}}
  \end{picture}
\begin{center}
(a)
\end{center}

  \begin{picture}(190,120)
    \put(30,0){\epsfxsize 180pt \epsfbox{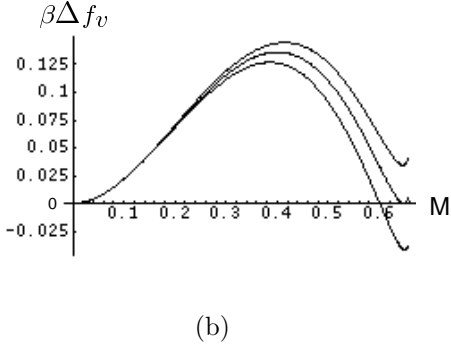}}
  \end{picture}
\begin{center}
(b)
\end{center}

  \caption{Typical behavior of free energy $f_v$ of (\ref{fv}) near
  phase transitions.
(a) $\beta fv(m,M)$ versus $m = \langle S_x \rangle$ for $\beta 
c_1=1, c_3=0$ near the second-order transition 
at $\beta c_2 = 0.166$. We set  $M = \langle J_{x\mu} \rangle$  at the value
on the transition point, $M = 0.999$.   
(b) $\beta \Delta f_v \equiv \beta (f_v(m,M) - f_v(0,0))$ versus 
$M $ for $\beta c_2=0.1, c_3=0$ near the first-order
transition at which $\beta c_1 = 0.678$, $m = m_c =0.989, M= M_c = 0.648$.
 The curves  are drawn along the line  $m = (m_c /M_c) M$ in the $(m,M)$
 plane.
 The three curves are for 
$\beta c_1 = 0.66, 0.678, 0.70$ from above.
}
\label{fig:fv}
\end{figure}

\newpage


\begin{figure} 
  \begin{picture}(200,110) 
    \put(10,0){\epsfxsize 175pt \epsfbox{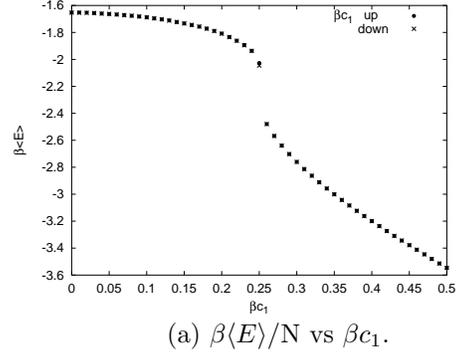}} 
  \end{picture} 
\vspace{-0.3cm}     
\begin{center}
(a) $\beta \langle E \rangle$/N vs $\beta c_1$.
\end{center}
\label{fig:Ufirst} 

  \begin{picture}(200,110) 
    \put(20,0){\epsfxsize 165pt \epsfbox{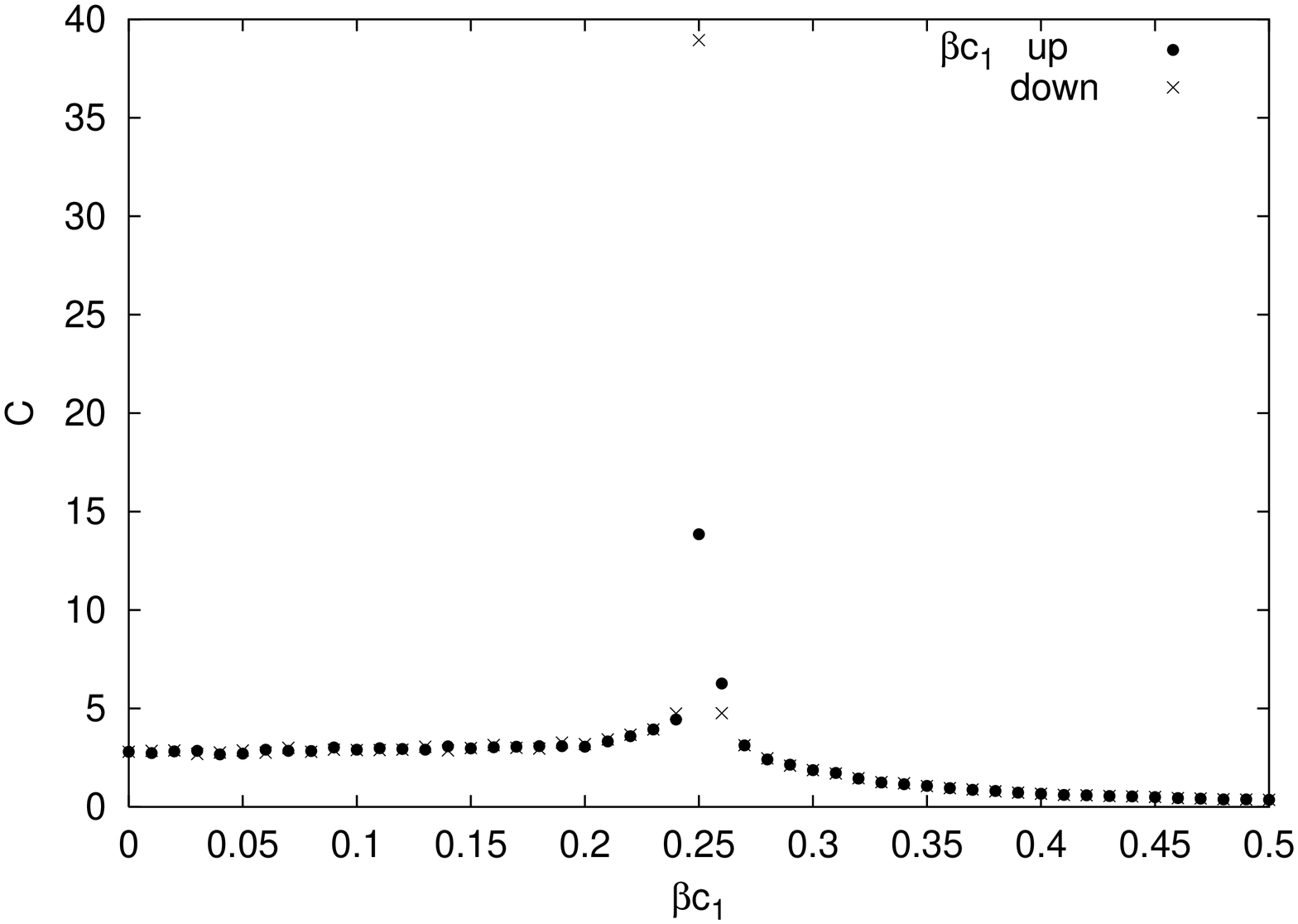}} 
  \end{picture} 
\vspace{-0.3cm}     
\begin{center}
(b) $C$/N vs $\beta c_1$.
\end{center}
 \caption{ Internal energy and specific heat per site versus 
 $\beta c_1$ for a typical first-order transition 
  ($ c_3=0.0, \beta c_2=0.7$). 
 In the measurement, $\beta c_1$ is increased first 
 (filled circles) and then
decreased (crosses).
} 
\label{fig:Cfirst} 
\end{figure} 


\begin{figure} 
  \begin{picture}(200,110) 
    \put(20,0){\epsfxsize 170pt \epsfbox{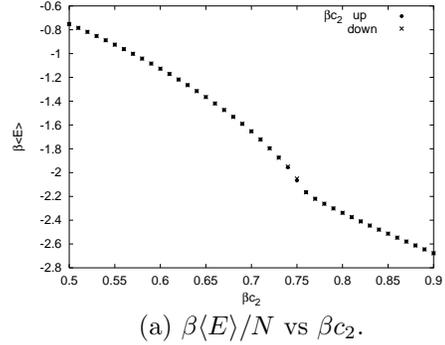}} 
  \end{picture}
\vspace{-0.3cm}   
\begin{center}
(a)  $\beta \langle E \rangle/N$ vs $\beta c_2$.
\end{center}
\label{fig:Usecond} 

  \begin{picture}(200,110) 
    \put(30,0){\epsfxsize 160pt \epsfbox{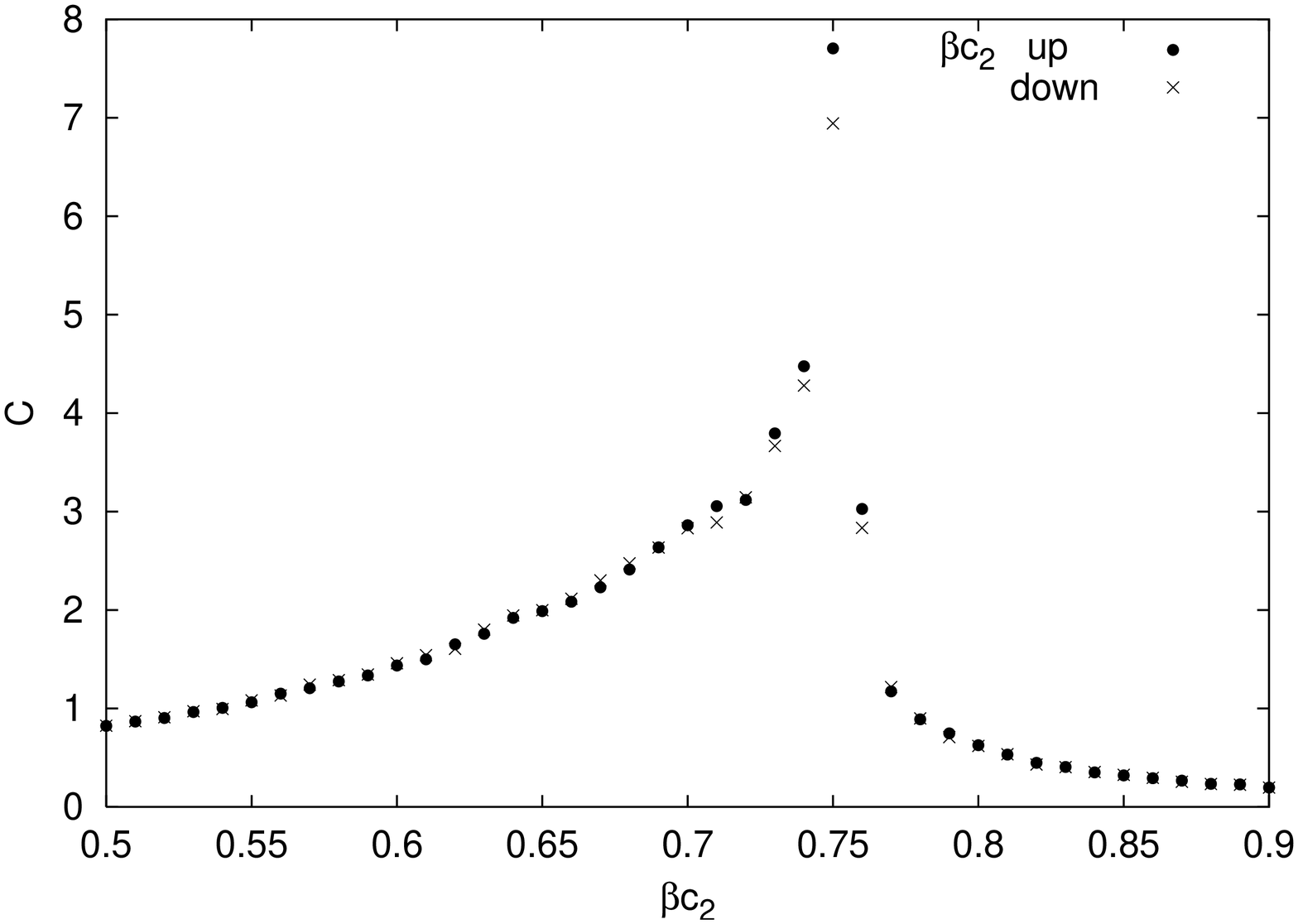}} 
  \end{picture} 
\vspace{-0.3cm}     
\begin{center}
(b)  $C/N$ vs $\beta c_2$.
\end{center}
  \caption{Internal energy and specific heat per site versus
   $\beta c_2$ for a typical second-order transition 
  ($\beta c_1=0.1, c_3 = 0.0$).
 In the measurement, $\beta c_2$ is increased first 
 (filled circles) and then decreased (crosses).
} 
\label{fig:Csecond} 
\end{figure}


\begin{figure} 
  \begin{picture}(200,330) 
    \put(20,0){\epsfxsize 230pt \epsfbox{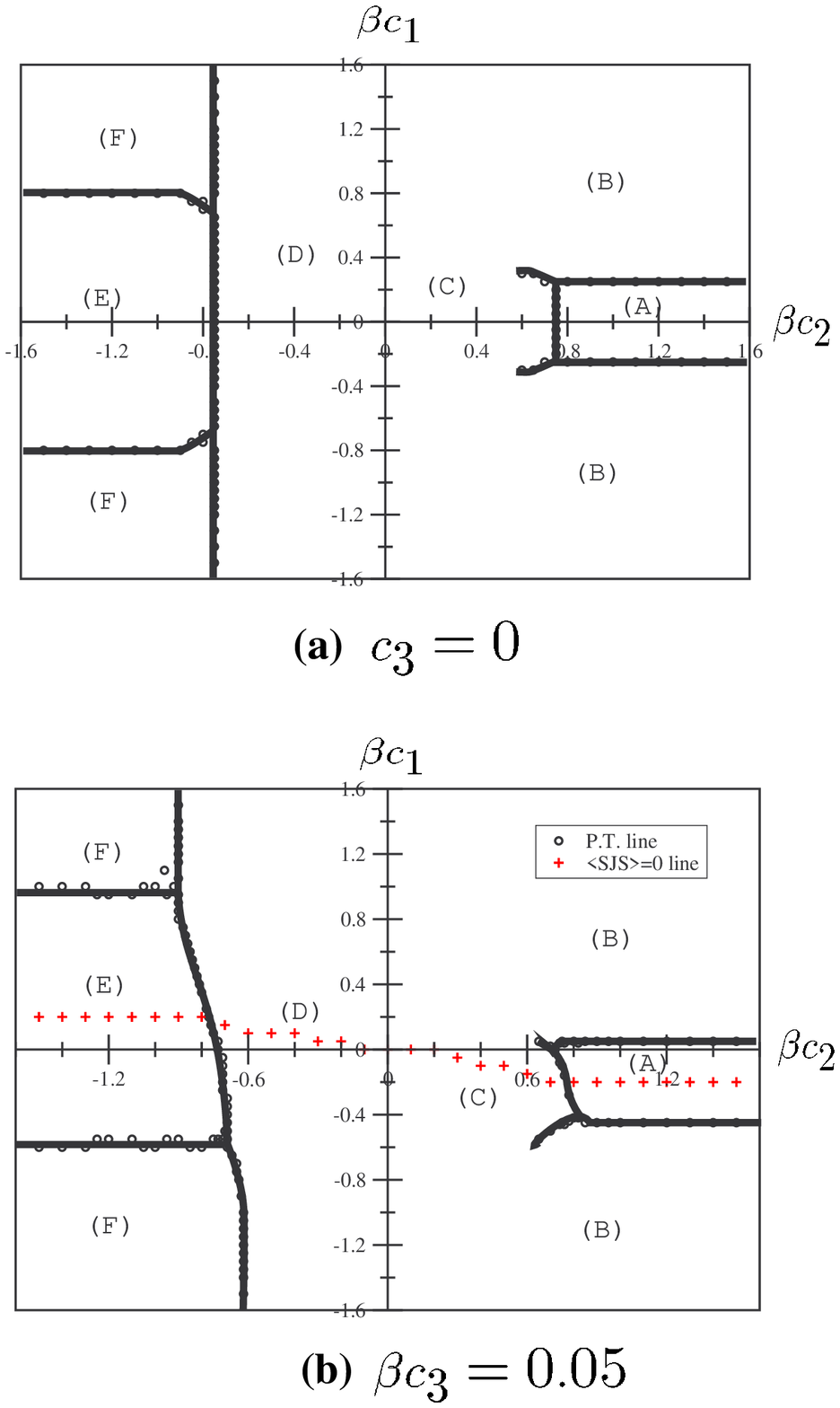}} 
  \end{picture} 
  \caption{Phase diagram including the extended regions $c_1,c_2 <0$
  for (a) $c_3=0$ and (b) $\beta c_3 = 0.05$.
  The phases D,E,F in $c_2 < 0$ are ``mirror images" of 
  C,B,A   in $c_2 > 0$, respectively. Crosses in (b) indicate the 
  points at which $j_{x\mu} = 0$. 
  (For the case (a), $j_{x\mu} = 0$ along the line $\beta c_1 = 0$.)
} 
\label{fig:phasediagram2} \
\end{figure} 

\vspace{8cm}


\begin{figure} 
  \begin{picture}(200,180) 
    \put(40,20){\epsfxsize 120pt \epsfbox{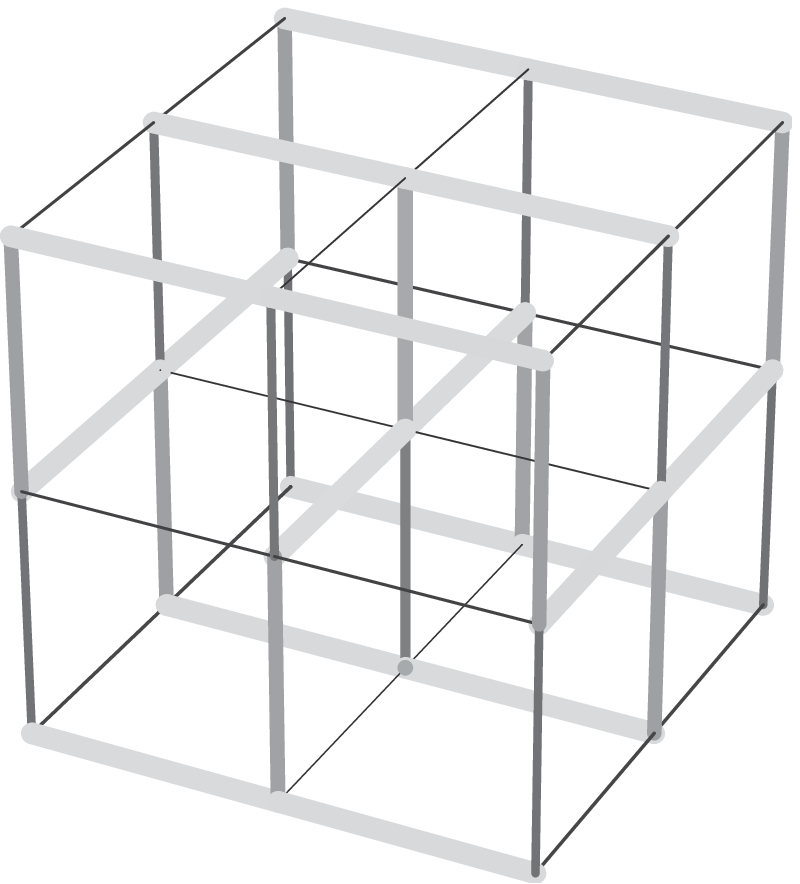}} 
    \put(160,25){\epsfxsize 20pt \epsfbox{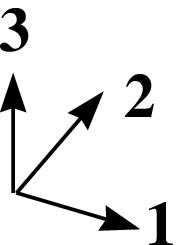}} 
  \end{picture} 
  \caption{Basic unit of the column-layer structure in the 
  ``Higgs" phase F in $c_2 < 0$.  The entire configuration of
  $j_{x\mu} \equiv \langle S_{x+\mu}J_{x\mu}S_x \rangle$ 
  is just the repetition of this 2x2x2 structure in every  direction. 
  The thickness of each segment represents the magnitude of $j_{x\mu}$. 
Each plane with a certain orientation 
(the 1-2 plane here)
has a one-dimensional columnic alignment of links with large 
(thick) $j_{x\mu}$ (say, in the 1-direction in the upper plane).
In the next (middle) plane, the direction of the columns with large 
$j_{x\mu}$ is rotated by 90 degrees 
(in the 2-direction), and so on.
Each pair of successive planes are weakly coupled by small (thin)
$j_{x\mu}$ (in the 3-direction).   
} 
\label{fig:layer} 
\end{figure}


\begin{figure} 
  \begin{picture}(200,200) 
    \put(30,0){\epsfxsize 140pt \epsfbox{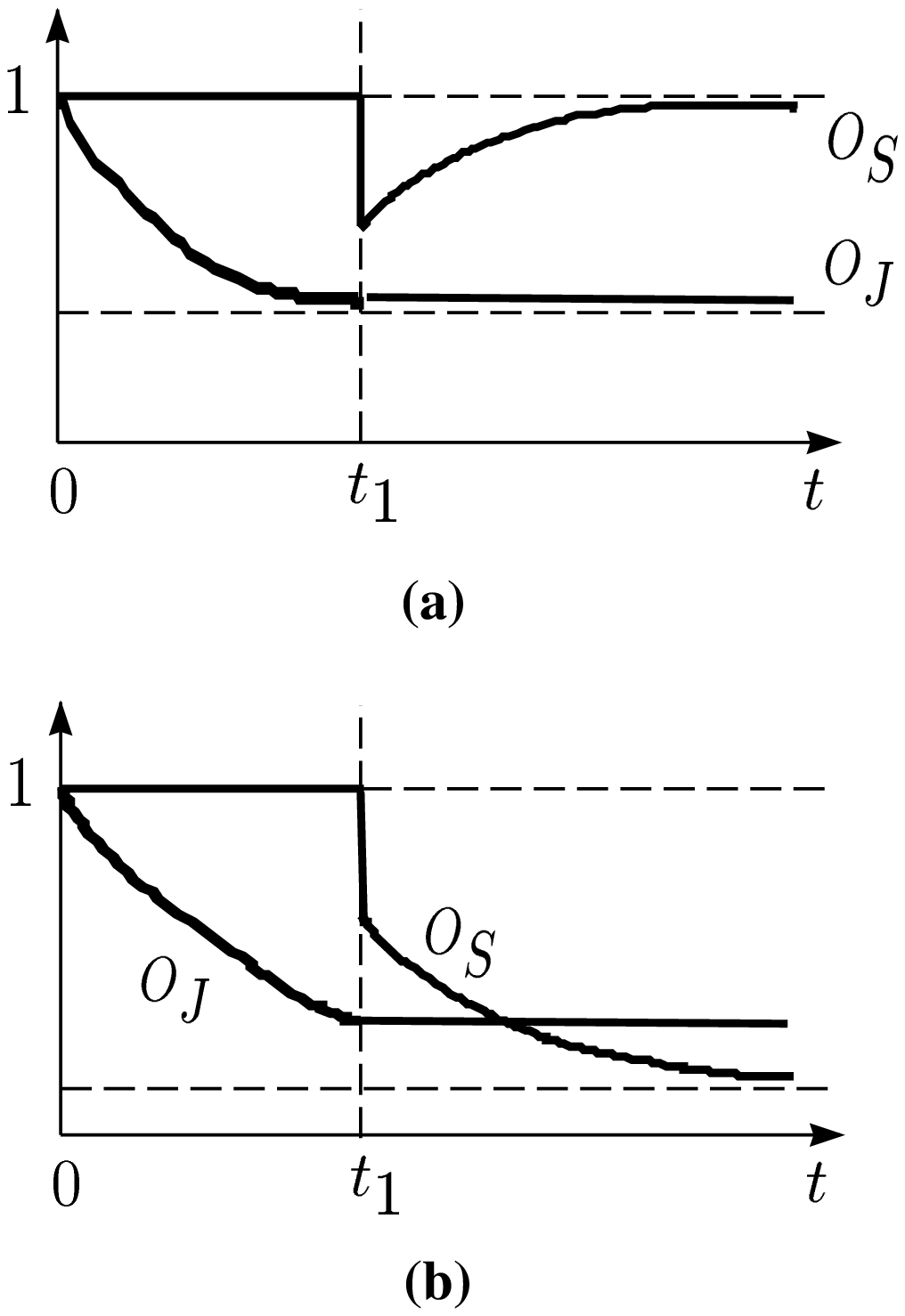}} 
  \end{picture} 
  \caption{Simulation of learning and recalling. For $0 < t < t_1$,
$S_x(t)$ is fixed to $S_x(0)$, and the system learns the 
pattern $S_x(0)$ by changing $J_{x\mu}(t)$.
At $t = t_1$, we disturb $S_x$ to $S_x(t_1)$ discontinuously,
where $S_x(t_1)$ is obtained by adding a certain amount of
random noise to $S_x(0)$. 
   For $t_1 < t$, 
   the system tries to recall $S_x(0)$ by changing $S_x(t)$.
   (We illustrate the case that $J_{x\mu}(t)$ for $t_1 < t$
   is fixed to $J_{x\mu}(t_1)$.) 
  (a) It succeeds to recall $S_x(0)$ with $S_x(t) \simeq S_x(0)$ and  
  $O_S \simeq 1$. (b) It fails to recall $S_x(0)$ with 
  $S_x(t) \neq S_x(0)$ and $O_S \neq 1$.
} 
\label{fig:typicalOs} 
\end{figure} 



\begin{figure}
  \begin{picture}(200,170) 
    \put(5,0){\epsfxsize 210pt \epsfbox{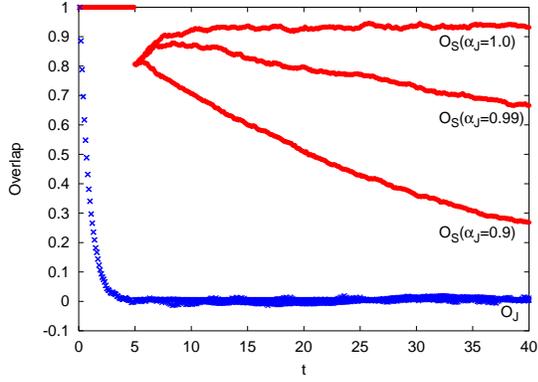}}
  \end{picture} 

\ \ \ \ \ \ \ \ \ \ (a)  $\beta c_1 = 1.0, c_2 = c_3 = 0$. 

  \begin{picture}(200,150) 
    \put(10,0){\epsfxsize 200pt \epsfbox{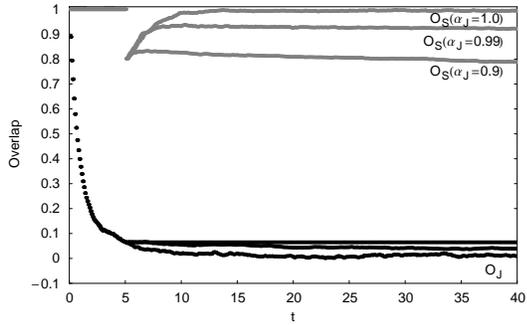}}
  \end{picture} 

\ \ \ \ \ \ \ \ \ \ (b)  $ \beta c_1 = 3.0, \beta c_2 = -1.0, c_3 = 0$. 

  \begin{picture}(200,150) 
    \put(10,0){\epsfxsize 200pt \epsfbox{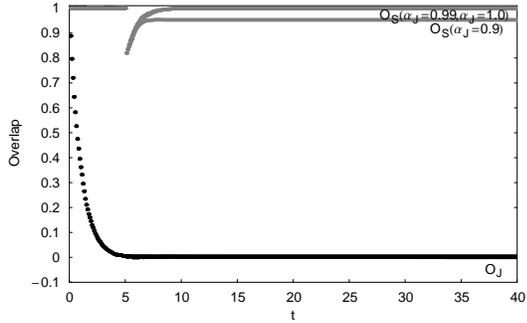}}
  \end{picture} 

\ \ \ \ \ \ \ \ \ \  (c)  
$ \beta c_1 = 3.0, \beta c_2 = 1.0, c_3 = 0$. 
\vspace{0.5cm}

  \caption{$O_S(t)$ and $O_J(t)$    in simultaneous
  time variations of $S_x$ and $J_{x\mu}$ for $t_1  < t$\ ($t_1 = 5$). 
  We choose $\alpha_J = 0.9$ for $t < t_1$, and 
  $\alpha_S = 0.9, \alpha_J = 1, 0.99, 0.9$ for $t_1 < t$.  
  The point of $c_i$ in the case (a) locates near 
  the Higgs-confinement phase boundary in MFT, 
  while the cases (b) and (c) locate  in the deep Higgs phase.
  $O_J$'s for three $\alpha_J$ in (a) and (b) are
  almost degenerate, while $O_J$'s in (b) are for 
  $\alpha_J = 1, 0.99, 0.9$ from the above, respectively.
  }
\label{fig:forget}
\end{figure} 


\begin{figure}
  \begin{picture}(200,130) 
    \put(30,0){\epsfxsize 170pt \epsfbox{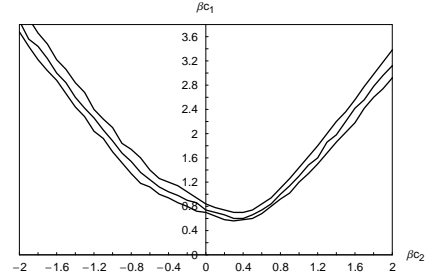}}
      \end{picture} 
      \begin{center}
\vspace{-0.2cm}
(a) $S_x(0)$ random;  $J_{x\mu}(0)$ random.
\end{center}
\vspace{-0.5cm}
  \begin{picture}(200,130) 
    \put(40,0){\epsfxsize 170pt \epsfbox{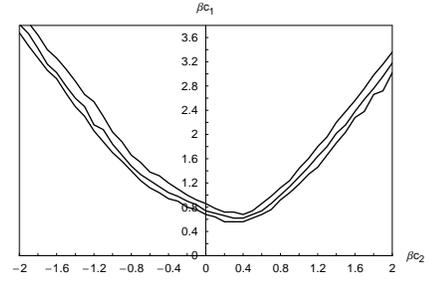}}
      \end{picture} 
      \begin{center}
\vspace{-0.2cm}
(b)  $S_x(0) = 1$;  $J_{x\mu}(0)$ random.
\end{center}
 \vspace{-0.5cm}
  \begin{picture}(200,130) 
    \put(40,0){\epsfxsize 170pt \epsfbox{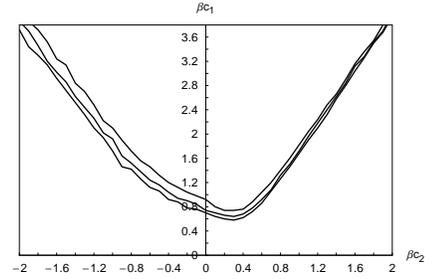}}
      \end{picture} 
      \begin{center}
\vspace{-0.2cm}
(c)  $S_x(0) = 1$;  $J_{x\mu}(0)=-1$.
\end{center}
 \vspace{-0.5cm}
  \begin{picture}(200,130) 
    \put(40,0){\epsfxsize 170pt \epsfbox{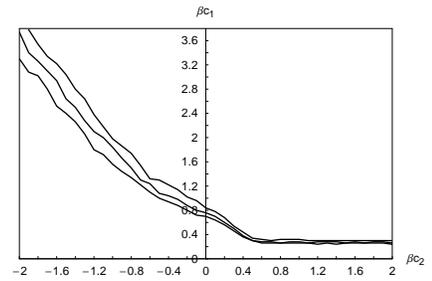}}
      \end{picture} 
      \begin{center}
\vspace{-0.2cm}
(d) $S_x(0)=1$;  $J_{x\mu}(0)=1$.
\end{center}

  \caption{Contour plots of $O_s$ at large $t$ 
  in the $\beta c_2- \beta c_1 $ 
  plane at $c_3 = 0$ for four different initial configurations (a)-(d).   
  Three curves in each figure are contours of $O_s = 0.9, 0.8, 0.7$
  from the above, respectively.  
The cases (a),(b),(c) look similarly each other,
  while the case (d) differs from them.}
\label{fig:Osinitial}
\end{figure}


\begin{figure} 
\begin{center} 
  \begin{picture}(200,110) 
    \put(10,0){\epsfxsize 160pt \epsfbox{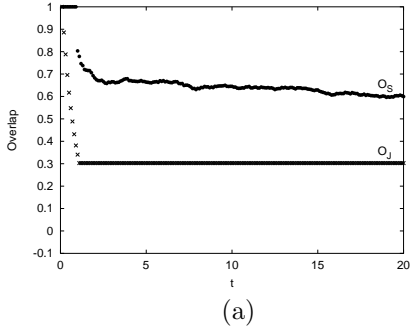}} 
  \end{picture}

(a)

   \vspace{0.5cm}
  \begin{picture}(200,110) 
    \put(10,0){\epsfxsize 160pt \epsfbox{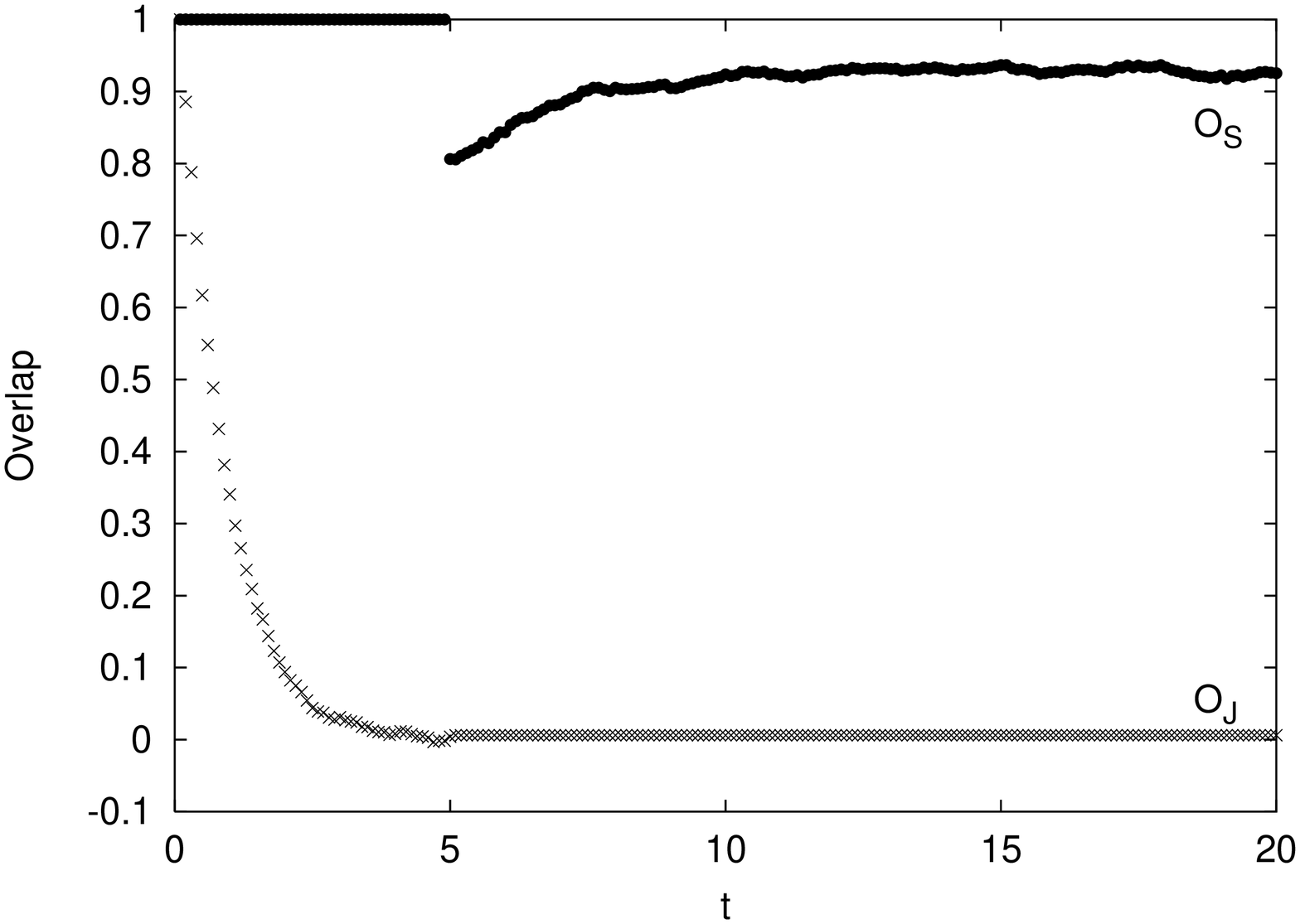}} 
  \end{picture} 

(b)
\end{center}
  \caption{ 
Overlaps $O_S, O_J$ for $\beta c_1=1.0,$\ $c_2=0.0,$\ $c_3=0.0$
with two different learning times $t_1$; 
(a) $t_1 = 1$ and (b) $t_1 = 5.$ The system fails to recall
$S_x(0)$  in (a) with a small $t_1$, while it succeeds
 in (b) with a suffciently large $t_1$.}
\label{fig:Ost1} 
\end{figure}


\begin{figure} 
  \begin{picture}(200,140) 
    \put(0,0){\epsfxsize 190pt \epsfbox{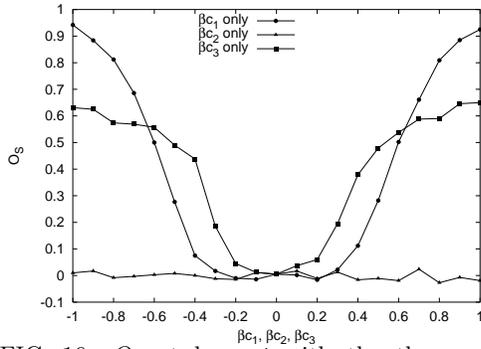}} 
  \end{picture} 
  \caption{$O_s$ at large $t$ with the three cases;
  $c_1$ only ($c_2=c_3=0$), $c_2$ only ($c_1=c_3=0$),
 and  $c_3$ only ($c_1=c_2=0$).} 
\label{fig:Osc123}
\end{figure} 

\newpage 

\begin{figure} 
  \begin{picture}(200,140) 
    \put(0,0){\epsfxsize 190pt \epsfbox{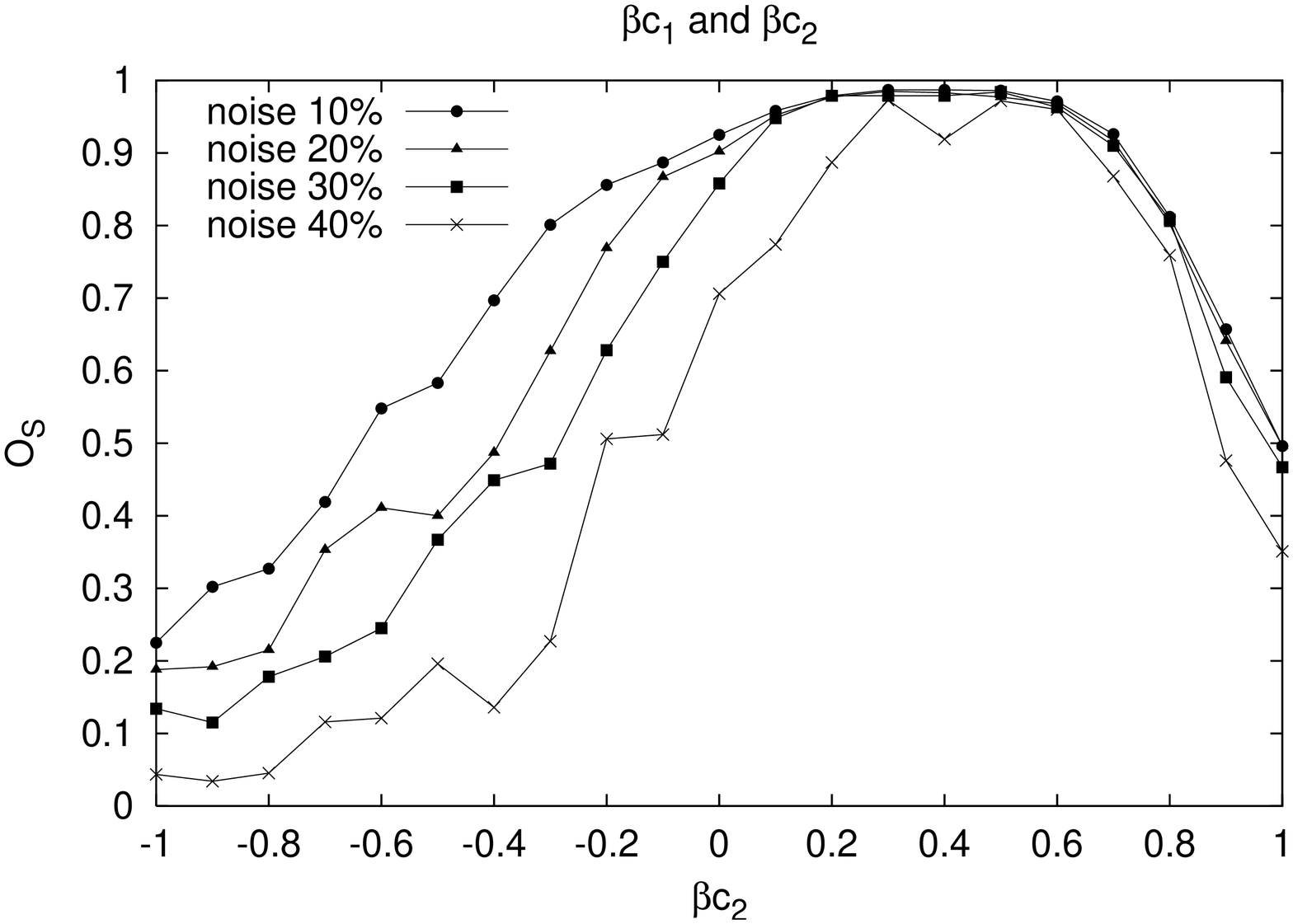}} 
  \end{picture}
\begin{center}
(a)
\end{center} 
\label{fig:kou}
 
  \begin{picture}(200,140) 
    \put(0,0){\epsfxsize 190pt \epsfbox{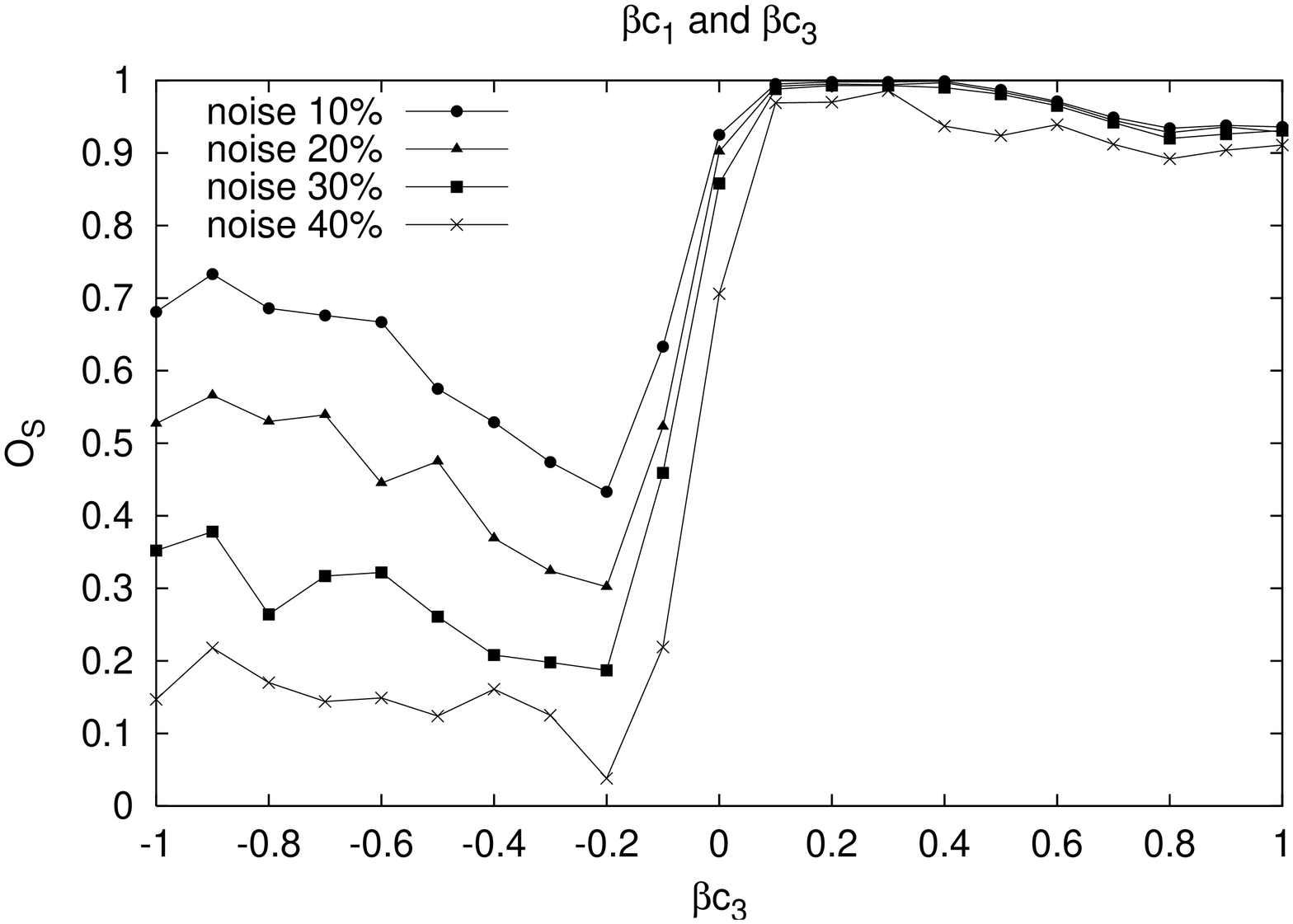}} 
  \end{picture} 
\begin{center}
(b)
\end{center}
  \caption{$O_S$ at sufficiently large time with several 
$S_x(t_1)$'s produced by adding various amounts of noises upon $S_x(0)$.
(a) $\beta c_1=1.0,  c_3=0.0$. 
(b) $\beta c_1 =1.0, c_2=0.0$.}
\label{fig:Osnoise}
\end{figure}


\begin{figure}
  \begin{picture}(200,140) 
    \put(0,0){\epsfxsize 220pt \epsfbox{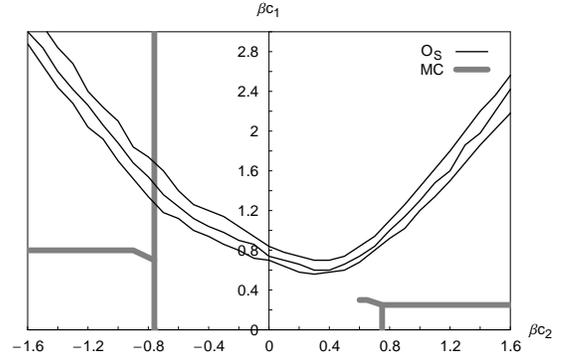}}
  \end{picture} 
  \vspace{0.5cm}
  \caption{Contour plot of $O_S$ in the $\beta c_2- \beta c_1$ plane
  at $c_3 = 0$. 
   Three curves are contours of $O_s = 0.9, 0.8, 0.7$
  from the above. The phase boundaries of MC simulations in
   Fig.\ref{fig:phasediagram2} are superposed.}
\label{fig:Oscontour}
\end{figure}


\begin{thebibliography}{99}
\bibitem{unwelcomeminima}
Here $\xi_i^\alpha$'s are assumed to be orthogonal each other, i.e.,
$N^{-1}\sum_i\xi_i^\alpha \xi_i^\beta = \delta_{\alpha \beta}$.
For $M >0.14 N$ there appear
unwelcome  energy minima other than $S_i = \xi^\alpha_i$.
\bibitem{kitaev2}
In Refs.\cite{kitaev}, the ordered phase of Z(2) gauge dynamics
is called the Higgs phase since the mass of gauge field is
massive there. This phase corresponds to Coulomb phase in the
present model, because $\langle S_i \rangle = 0$.
The name Higgs phase is reserved for the phase with 
$\langle S_i \rangle \neq 0$ [see (\ref{tab:phases})].
\bibitem{jij0}
More faithful generalization of the Tanaka model\cite{tanaka} 
may lead us to the variable
$J_{ij} = -1, 0, 1$. (The constraint $\sum_j J_{ij}= 1$ is to be 
irrelevant since the signature of $J_{ij}$ is indefinite here.) 
Such a case is discussed in \cite{matsui}
as the Model II on a 3D lattice. We made also some MC simulations
for this case with the energy given in Sect.IIIA, 
which predict phase diagrams and  
a column structure similar to the case of Z(2) gauge variables
presented in Sect.IVB and IVC.
\bibitem{difference}
Although $S_x$ and $J_{x\mu}$ are not continuous but discrete
variables, $\Delta E$ is proportionalto the formal derivative
like $\partial E /\partial S_x$.
\bibitem{fluxless}
A simple way to see this is to make an expansion in powers of
$c_1$ (assuming it is small). After the $S_x$ summation,
a plaquette term like the $c_2$ term is generated at $O(c_1^4)$
with a coefficient favoring the fluxless states.
\bibitem{symmetry}
The sites of 3D lattice are partitioned into odd sites
and even sites according to the parity 
$(-)^{x_1 + x_2 + x_3}$. By making changes of variables
$S_x \rightarrow S_x' \equiv (-)^{x_1 + x_2 + x_3} S_x$
the energy changes the signatures of the $c_1$ and $c_3$
terms. 
$Z$ must be invariant under  this changes of variables,
which proves (\ref{symmetry}).
\bibitem{ajless1}
We also drew some figures using 
$O_S$ at a suitable large time which are calculated 
with $\alpha_J(t_1<t) < 1$. 
As expected from Fig.\ref{fig:forget}, 
they deviate systematically from  
Fig.\ref{fig:Osinitial}-\ref{fig:Oscontour}.
However, they look qualitatively similar each other, 
and we found no significantly new phenomena.
\end{thebibliography}
\end{document}